\documentclass[12pt,preprint]{aastex}
\begin{document}
\def\hh{\, h^{-1}}
\newcommand{\wth}{$w(\theta)$}
\newcommand{\xir}{$\xi(r)$}
\newcommand{\Lya}{Ly$\alpha$}
\newcommand{\Lyb}{Lyman~$\beta$}
\newcommand{\Hb}{H$\beta$}
\newcommand{\HI}{H{\sc I}}
\newcommand{\msun}{M$_{\odot}$}
\newcommand{\sfr}{M$_{\odot}$ yr$^{-1}$}
\newcommand{\sfrd}{M$_{\odot}$ yr$^{-1}$ Mpc$^{-3}$}
\newcommand{\cld}{erg s$^{-1}$ Hz$^{-1}$ Mpc$^{-3}$}
\newcommand{\dnsty}{$h^{-3}$Mpc$^3$}
\newcommand{\za}{$z_{\rm abs}$}
\newcommand{\ze}{$z_{\rm em}$}
\newcommand{\cmtwo}{cm$^{-2}$}
\newcommand{\nhi}{$N$(H$^0$)}
\newcommand{\degpoint}{\mbox{$^\circ\mskip-7.0mu \,$$~$}}
\newcommand{\halpha}{\mbox{H$\alpha$}}
\newcommand{\hbeta}{\mbox{H$\beta$}}
\newcommand{\hgamma}{\mbox{H$\gamma$}}
\newcommand{\kms}{\,km~s$^{-1}$}      % note leading thinspace
\newcommand{\minpoint}{\mbox{$'\mskip-4.7mu \mskip0.8mu \,$$~$}}
\newcommand{\mv}{\mbox{$m_{_V}$}}
\newcommand{\Mv}{\mbox{$M_{_V}$}}
\newcommand{\peryr}{\mbox{$\>\rm yr^{-1}$}}
\newcommand{\secpoint}{\mbox{$''\mskip-7.6mu \\,$}}
\newcommand{\sqdeg}{\mbox{${\rm deg}^2$}}
\newcommand{\squig}{\sim\!\!}
\newcommand{\subsun}{\mbox{$_{\twelvesy\odot}$}}
\newcommand{\et}{{\it et al.}~}
\newcommand{\er}[2]{$_{-#1}^{+#2}$}
\def\h50{\, h_{50}^{-1}}
\def\hbl{km~s$^{-1}$~Mpc$^{-1}$}
\def\ltsima{$\; \buildrel < \over \sim \;$}
\def\gtsima{$\; \buildrel > \over \sim \;$}
\def\spose#1{\hbox to 0pt{#1\hss}}
\def\simlt{\mathrel{\spose{\lower 3pt\hbox{$\mathchar"218$}}
     \raise 2.0pt\hbox{$\mathchar"13C$}}}
\def\simgt{\mathrel{\spose{\lower 3pt\hbox{$\mathchar"218$}}
     \raise 2.0pt\hbox{$\mathchar"13E$}}}
\def\arcs{$''~$}
\def\arcm{$'~$}
\newcommand{\wu}{$U_{300}$}
\newcommand{\wb}{$B_{435}$}
\newcommand{\wv}{$V_{606}$}
\newcommand{\wi}{$i_{775}$}
\newcommand{\wz}{$z_{850}$}
\newcommand{\hmpc}{$h^{-1}$Mpc}
\newcommand{\um}{$\mu$m}
\title{Spectroscopic Observations of Lyman--Break Galaxies \\
at Redshift $\sim$ 4, 5 and 6 in the GOODS--South Field\altaffilmark{1}}

\author{\sc E. Vanzella\altaffilmark{2}, 
            M. Giavalisco\altaffilmark{3}, 
            M. Dickinson\altaffilmark{4},
	    S. Cristiani\altaffilmark{2,10},
	    M. Nonino\altaffilmark{2}, \\
	    H. Kuntschner\altaffilmark{5},
	    P. Popesso\altaffilmark{6},
	    P. Rosati\altaffilmark{6},
            A. Renzini\altaffilmark{7},
            D. Stern\altaffilmark{8},\\
	    C. Cesarsky\altaffilmark{6},
            H. C. Ferguson\altaffilmark{9},
            R.A.E. Fosbury\altaffilmark{5},\\
	    and the GOODS Team} 

\affil{$^{2}$INAF -- Osservatorio Astronomico di Trieste, via G.B. Tiepolo 11,
  40131 Trieste, Italy}
\affil{$^{3}$Astronomy Department, University of Massachusetts, Amherst MA
  01003, USA} 
\affil{$^{4}$NOAO, PO Box 26732, Tucson, AZ 85726, USA}
\affil{$^{5}$ST--EFC, Karl Schwarzschild Strasse 2, 85748, Garching, Germany}
\affil{$^{6}$ESO, Karl Schwarzschild Strasse 2, 85748, Garching, Germany}
\affil{$^{7}$INAF -- Osservatorio Astronomico di Padova, Vicolo
  dell'Osservatorio 5, 35122, Padova, Italy}
\affil{$^{8}$JPL, California Institute of Technology, Mail Stop 169-527,
  Pasadena, CA 91109} 
\affil{$^{9}$STScI, 3700 San Martin Dr., Baltimore, MD 21218, USA}
%\affil{$^{10}$Department of Astronomy, University of California Berkeley,
%  Berkeley, CA 94720}
\affil{$^{10}$ INFN, National Institute of Nuclear Physics, via Valerio 2,
  I-34127 Trieste, ITALY}

\altaffiltext{1}{Based on observations made at the European Southern
  Observatory Very Large Telescope, Paranal, Chile (ESO programme 170.A-0788
  The Great Observatories Origins Deep Survey: ESO Public Observations of the
  SST Legacy / HST Treasury / Chandra Deep Field South). Also based on
  observations obtained with the NASA/ESA {\it Hubble Space Telescope}
  obtained at the Space Telescope Science Institute, which is operated by the
  Association of Universities for Research in Astronomy, Inc. (AURA) under
  NASA contract NAS 5-26555.}

\begin{abstract} %{\bf \today}

We report on observations of Lyman--break galaxies (LBGs) selected
from the Great Observatories Origins Deep Survey (GOODS) at mean
redshifts $z\sim 4$, 5 and 6 (\wb--, \wv--\ and \wi--band dropouts,
respectively), obtained with the red--sensitive FORS2 spectrograph
at the ESO VLT.  This program has yielded spectroscopic identifications
for 114 galaxies ($\sim$ 60\% of the targeted sample), of which 51
are at $z\sim 4$, 31 at $z\sim 5$, and 32 at $z\sim 6$.  We demonstrate
that the adopted selection criteria are effective, identifying
galaxies at the expected redshift with minimal foreground contamination.
Of the 10\% interlopers, 83\% turn out to be
Galactic stars.  Once selection effects are properly accounted for,
the rest--frame UV spectra of the higher--redshift LBGs appear to
be similar to their counterparts at $z\sim 3$.  As at
$z\sim 3$, LBGs at $z\sim 4$ and $z\sim 5$ are observed with \Lya\
both in emission and in absorption; when in absorption, strong
interstellar lines are also observed in the spectra. The stacked
spectra of \Lya\ absorbers and emitters also show that the former
have redder UV spectra and stronger but narrower interstellar lines,
a fact also observed at $z\sim 2$ and $3$. At $z\sim 6$,
sensitivity issues bias our sample towards galaxies with \Lya\ in
emission; nevertheless, these spectra appear to be similar to their
lower--redshift counterparts. As in other studies at similar
redshifts, we find clear evidence that brighter LBGs tend to have
weaker \Lya\ emission lines. At fixed rest--frame UV luminosity, the
equivalent width of the \Lya\ emission line is larger at higher
redshifts. At all redshifts where the measurements can be reliably
made, the redshift of the \Lya\ emission line turns out to be larger
than that of the interstellar absorption lines, with a median
velocity difference $\Delta V\sim 400$ \kms\ at $z\sim$ 4 and 5,
consistent with results at lower redshifts.  This shows
that powerful, large--scale winds are common at high redshift.  In
general, there is no strong correlation between the morphology of
the UV light and the spectroscopic properties.  However, galaxies
with deep interstellar absorption lines and strong \Lya\ absorption
appear to be more diffuse than galaxies with \Lya\ in emission.

\end{abstract}
\keywords{cosmology: observations --- galaxies: formation --- galaxies: 
evolution --- galaxies: distances and redshifts}

\section{Introduction}

The study of galaxies at high redshift is crucial for understanding
the formation of the Hubble sequence, the growth of visible structures
in the Universe and the processes leading to the reionization of
the intergalactic hydrogen at the end of the Dark Ages. In the past
decade, the empirical investigation of galaxies at high redshifts
(i.e., $z>1.5$) has made rapid progress thanks to advances in
telescopes and instrumentation and to the development of optimized
selection techniques based on the observed colors of galaxies through
either broad or narrow passbands. Color--selection criteria, which
are designed to target galaxies with a range of spectral
energy distributions (SEDs) within a targeted redshift window, are
generally very efficient and allow one to build large samples with
reasonably well controlled systematics \citep[e.g.,][]{steidel99,
dad04, vandokkum03, tanigu05}, suitable for a broad range of studies,
both statistical in character or based on the properties of the
individual sources.

Among the various types of galaxies at high redshifts identified
by color selection, the Lyman--break galaxies \citep[LBGs; e.g.,][for
a review, see Giavalisco 2002]{guha90,steidel03, giava04b} are the
best studied and their samples are the largest both from a statistical
point of view, and in terms of the the cosmic time covered (reaching
back to less than one billion years after the Big Bang).  The reason
is mostly practical: since these galaxies are selected on the basis
of luminous rest--frame ultraviolet (UV) emission, which at redshifts
$2.5 \simlt z \simlt 6$ is redshifted into the optical and
near--infrared windows, the observations are among the easiest to
carry out, taking advantage of very sensitive instrumentation with
large areal coverage.

LBGs at redshift $z\sim 3$ have been intensively studied with both
very large \citep{steidel03} and deep samples \citep{papovich01},
including high--quality spectra for more than a thousand galaxies.
Current surveys at $z\sim 3$ provide the largest data set to study
the properties of galaxies during a relatively early phase of galaxy
evolution ($z\sim 3$ corresponds to when the Universe was $\sim
20$\% of its current age), at least for one spectral type, namely
star--forming galaxies with moderate dust obscuration. A number of
follow--up studies have been carried out following the discovery
of these galaxies \citep{steidel96a, steidel96b}, including studies
of their morphology and size \citep{giava96, papo03, ravi06, law07,
lotz06, ferg04}, their ages and stellar masses \citep{papo00,
shapley00, dick03a}, their chemical evolution \citep{pettini00,
shapley03}, and of their clustering properties \citep{giava98,
adel98, giava01, adel03}.

At higher redshifts, LBGs appear fainter, the observations require
higher sensitivity and the samples are still relatively small. As
a consequence, the properties of the most--distant LBGs are less well
characterized.  Initial studies of the spectral properties and
luminosity function have been carried out at $z\sim 4$ \citep{steidel99}
and $z\sim 5$ \citep{madau98} based on fairly small samples. More
recently, larger samples at $z>4$ have been gathered, from both
ground-- and space--based observatories \citep{shima05, iwata03,
giava04b, dick04, bunk04, bouwens06, bouwens07}; these samples,
however, are exclusively photometric ones, with small numbers of
spectroscopic identifications.  Such studies have 
investigated a wide spectrum of statistical properties of LBGs at
$z>4$ and up to $z\sim 7$, such as spatial clustering, morphology,
UV luminosity function, stellar mass and properties of the \Lya\
line \citep[see, for example,][]{hamana06, lee06, ravi06, lotz06,
giava04b, ferg04, ouchi05, bouwens07, ando06, ando07, yan06}.
However, these results are based on the assumption that the spectral
properties of LBGs at $z \simgt 4$ are the same as at $z\sim 3$. It is
reasonable to expect that the higher--redshift samples should bear
a similarity to those at $z\sim 3$, since LBG color selection at
all redshifts are tuned to select galaxies with a similar rest-frame
UV SED.  However, without spectroscopic information, it is impossible
to know if evolutionary effects are introducing systematic biases in
the observed statistics. For example, if the distribution of surface
brightness, UV SED or \Lya\ emission line properties evolve with
redshift, this will affect the redshift distribution and the
completeness of the samples, which in turn will bias derived
properties such as the LBG spatial clustering, luminosity function,
and the evolution of these quantities.\footnote{Note that a simple
analysis of the {\it observed} colors or sizes is not sufficient
to establish the presence of evolutionary effects since it is not
possible to separately measure the distribution functions of color,
size and luminosity.  See discussion in \cite{reddy08}.}

In this paper we present results from a program of spectroscopic
follow--up of LBGs at redshift $z>4$ selected from optical
images obtained with the Advanced Camera for Surveys (ACS; Ford et
al.\ 1993) on board the {\it Hubble Space Telescope} ({\it HST})
in the four passbands, \wb, \wv, \wi\ and \wz, as part of the Great
Observatories Origins Deep Survey, or GOODS \citep[for an overview
of the GOODS project, see][]{renz02, dick03b, giava04a}. The spectra
have been obtained at the ESO VLT with the FORS2 spectrograph. 
%Part of 
The data from this program, specifically all the spectra of
galaxies in the redshift range 0.5-6.3, have already been released and
described in previous papers (Vanzella et al.\ 2005, 2006, 2008).
Here we focus on a sample of LBGs, which has yielded 114 spectroscopic
identifications in the redshift interval $3.1 - 6.3$. Re-analyzing 
the whole LBG sample, we introduce very few differences 
(mainly in the quality redshift) 
to respect the previous global release (Vanzella et al. \ 2008), improvements
that have been marked in the reported list of the present work.
Currently, it represents one of the largest and most homogeneously--selected
spectroscopic samples in this redshift range.

Throughout this paper magnitudes are in the AB scale (Oke 1974),
and the world model, when needed, is a flat universe with density
parameters $\Omega_m=0.27$, $\Omega_{\Lambda}=0.73$ and Hubble
constant $H_0=73$ \hbl.

\section{Data and Sample Selection}

\subsection{ACS Images and Source Catalogs}

We have selected samples of LBGs at mean redshift $z\sim 4$, 5 and
6 (in the following referred to as \wb--, \wv--, and \wi--band
dropouts, respectively) from the latest version (v2.0) of the GOODS
images, obtained with the ACS on {\it HST}.  The v2.0 mosaics are
nearly identical in shape and size to the v1.0 ones. They cover the
two GOODS fields, the northern one encompassing the {\it Hubble}
Deep Field North (HDF--N) and the southern one located at the center
of the {\it Chandra} Deep Field South (CDF--S).  Each subtend an
area of approximately $10\times 17$ arcmin on the sky, for a total
areal coverage of about 0.1 square degrees.  As with v1.0, the v2.0
images consist of two sets of mosaics observed in the \wb, \wv,
\wi\ and \wz\ filters.  The depth of the \wv, \wi\ and \wz\ mosaics,
however, has been increased over version v1.0 by including additional
observations taken during the continuation of the original GOODS
survey for high-redshift Type~Ia supernovae \citep{riess04, riess05}.
Since these additional data were obtained using the same observational
strategy as the original ACS program (e.g., the same Phase-II files
were used to carry out the observations), integrating them into the
existing mosaics has been straightforward and has resulted in
doubling the original exposure time in the \wz\ band as well as a
more modest depth increase in the other bands.\footnote{The v2.0
exposure times in the \wb\ , \wv\ , \wi\ and \wz\ bands are 7200,
5450, 7028 and 18232 seconds, respectively.  Details of the ACS
observations, as well as major features of the GOODS project, can
be found in \cite{giava04a}; additional information about the latest
v2.0 release of the GOODS ACS images and source catalogs can be
found at {\tt www.stsci.edu/science/goods/v2.0} and will be described
in detail in an upcoming paper (Giavalisco et al., in prep.).}

\subsection{Photometric Samples of Lyman--Break Galaxies}

We have selected samples of LBGs using color criteria very similar
to those presented by Giavalisco et al.\ (2004b; G04b hereafter),
with some minor modifications applied to the definition of \wb--band
dropouts to explore the redshift distribution of galaxies near the
border of that color--color selection window.  The exact locations
of such windows balance the competing desires of completeness and
reliability.  Windows are designed to include as complete a sample
of target galaxies as possible given the dispersion of observed
colors --- due both to both observational scatter and the intrinsic
dispersion in galaxy UV SEDs (e.g., related to varying dust content,
ages, metallicities, \Lya\ equivalent widths, etc.).  On the other
hand, windows are designed to avoid significant numbers of galaxies
at redshifts outside (usually lower than) the targeted one.

In the present work, \wb--band dropouts are defined as objects that
satisfy the color equations: $$(B_{450}-V_{606})\ge 1.1 +
(V_{606}-z_{850})~~~\wedge~~~ (B_{450}-V_{606})\ge 1.1~~~\wedge~~~
(V_{606}-z_{850})\le 1.6,\eqno(1.1)$$ where $\wedge$ and $\vee$ are
the logical AND and OR operators.  These criteria extend the selection
of candidates to slightly bluer (\wb-\wv) and redder (\wv-\wz)
colors than those in G04b. As can be seen in Figure~\ref{Bdrop},
which shows the selection windows corresponding to both sets of
color equations, the sample selected with the new criteria (solid
line) fully includes the one selected with the G04b criteria (dashed
line).  We have decided to use these more general criteria to define
the sample of \wb--band dropouts, which is the largest among the
three LBG samples targeted for the spectroscopic observations, to
explore both changes in the low--end of the targeted redshift range
and contamination rates from low--redshift interlopers.

The definitions of the color equations of \wv--band and \wi--band
dropouts are unchanged from those used in G04b and \cite{dick04},
and are given by the color equations $$[(V_{606}-i_{775})>1.5+0.9\times
(i_{775}-z_{850})]~~~\vee$$
$$\vee~~~[(V_{606}-i_{775})>2.0]~~~\wedge~~~(V_{606}-i_{775})\ge
1.2~~~\wedge~~~(i_{775}-z_{850})\le 1.3$$ $$\wedge
~~~[(S/N)_{B}<2]\eqno(1.2)$$ and $$(i_{775}-z_{850})>1.3 ~~~\wedge
~~~[(S/N)_{B}<2] ~~~\vee~~~[(S/N)_{V}<2],\eqno(1.3)$$ respectively
(see Figure~\ref{Vdrop} for a collapsed representation of the
selection windows for \wv-- and \wi\--band dropouts).  For all three
selection criteria above, when the isophotal $S/N$ in a given band
is less than one, limits on the colors have been calculated using the
1$\sigma$ error on the isophotal magnitude.

We have restricted the photometric samples to galaxies with isophotal
$S/N\ge 5$ in the \wz\ band, and we have visually inspected each
candidate, removing sources that were deemed artifacts.  In addition,
we have estimated the number of spurious detections using counts
of negative sources detected in the same data set.  Together, these
amount to a negligible number of spurious sources for the \wb-- and
\wv--dropout samples, and $\approx 12$\% for the \wi--dropouts.  We
have also eliminated all sources with stellar morphology down to
apparent magnitude \wz$\sim 26$, i.e., where such a morphological
classification is reliable.  This accounts for an additional $3.1$\%,
$8.3$\% and $4.6$\% of the \wb--, \wv-- and \wi--dropout samples,
respectively.  While this procedure biases our samples against LBGs
(and high--redshift quasars) that are unresolved by ACS, it minimizes
contamination from Galactic stars.  Note that we have spectroscopically
observed a few point-like sources that obey the dropout selections
in order to verify that such sources are indeed Galactic.  In
practice, these cullings of the dropout samples result in negligible
changes to the spectroscopic samples and to key measured quantities,
such as the specific luminosity density.

Down to \wz$\le 26.5$, roughly the $50$\% completeness limit for
unresolved sources, the culled samples include 1544, 490 and 213
\wb--, \wv--and \wi--band dropouts, respectively. With a
survey area of 316 arcmin$^2$, this corresponds to surface density
$\Sigma=4.89\pm 0.12$, $1.55\pm 0.07$ and $0.67\pm 0.05$ galaxies
per arcmin$^2$ for the three types of dropouts, respectively.  Error
bars simply reflect Poisson fluctuations.

We note that while the \wv-- and \wi\--band dropout samples are
mutually exclusive (i.e., $i_{775}-z_{850}\leq 1.3$ vs.
$i_{775}-z_{850}>1.3$, respectively), the intersection of the \wb--
and \wv\--band dropout samples may be non-zero.  However, in this
latter case, no sources in common have been found down to the \wz$\le
26.5$, and only one galaxy satisfies both criteria when the magnitude
limit is extended down to \wz$\le 27.5$ (i.e., GDS~J033245.88-274326.3).

We have used Monte Carlo simulations to estimate the redshift
distribution function of our LBG samples and compared the results
to observations.  The technique is the same as that used in G04b
and consists of generating artificial LBGs distributed
over a large redshift range (we used $2.5\le z\le 8$) with assumed
distribution functions for UV luminosity (we used a flat distribution,
discussed below), SED, morphology and size.  We adjusted the input
SED and size distribution functions by requiring that the distribution
functions recovered from the simulations match the $z\sim 4$ observed
sample, the largest of the three GOODS samples.  In this way, both
simulations and observations are subject to similar incompleteness,
photometric errors (in flux and color), blending, and other measurement
errors.  The model SED used for the simulations is based on a
synthetic spectrum of a continuously star--forming galaxy with age
$10^8$ yr, Salpeter IMF and solar metallicity (Bruzual \& Charlot
2000).  We reddened it with the starburst extinction law (Calzetti
2000) and $E(B-V)$ randomly extracted from a Gaussian distribution
with $\mu_{\rm E(B-V)}=0.15$ and $\sigma_{\rm E(B-V)}=0.15$.  In
other words, the dispersion of the LBG UV SEDs is modeled as only
due to the dispersion in the amount of obscuration for the same
unobscured SED, neglecting the effects of age and metallicity of
the stellar populations.  This is obviously a crude approximation,
but, thanks to the strong degeneracy between age, obscuration and
metallicity on the broad--band UV colors of star--forming galaxies,
it is adequate here since we are only interested in measuring the
selection effects due to the specifics of the observations. For the
cosmic opacity, we have adopted the Madau (1995) prescription,
extrapolated to higher redshifts when necessary. To model
the dispersion of the morphologies of the galaxies we have used an
equal number of $r^{1/4}$ and exponential profiles with random
orientation, and size extracted from a log--normal distribution
function \citep[see][]{ferg04}. We found the average redshift
and standard deviation of the redshift distribution to be $z_B=3.78$
and ${\sigma}_B=0.34$ for the \wb--dropout sample, $z_V=4.92$ and
${\sigma}_V=0.33$ for the \wv--dropout sample, and $z_i=5.74$ and
${\sigma}_i=0.36$ for the \wi--dropout sample.

\subsection{The Spectroscopic Sample}

We have selected a sample of 202 LBGs from the three samples defined
above as primary targets of the FORS2 spectroscopic observations.
While the criteria to include a galaxy in the target list were
mostly based on its apparent magnitude, as we detail below, we did
not set a strict flux limit for the spectroscopic sample in these
initial high-redshift LBG spectroscopic studies.  This allowed us
to empirically assess how often the presence of \Lya\ emission
allows the measurement of the redshift of galaxies which are too
faint for absorption spectroscopy.

Targets were assigned slits in the FORS2 multi--object spectroscopic
masks according to an algorithm in which two competing factors
combine to maximize (i) the number of targets and (ii) the likelihood
of success, under the assumption that brighter targets are more
likely to result in successful identifications.  In practice, while
brighter galaxies were more likely to be assigned a slit, (slightly)
fainter targets could still win the competition if their coordinates
allowed a larger total number of targets on a given mask.  Relatively
faint targets in close proximity to brighter one were also assigned
a slit if their inclusion could be made without penalty.  Where
possible, we assigned faint targets to multiple masks.

When the number of available slits in a mask exceeded that of
available targets, we populated the remaining slits with ``filler''
targets selected to test target selection criteria and to identify
lower--redshift galaxies in the range $z \sim 1-2$ \citep[for a summary
of the global target selection of the FORS2 campaign, see][]{vanz06}.
In particular, we selected some filler targets using LBG criteria
that extended the primary \wb-- and \wv--band dropout selection
criteria to galaxies with less pronounced ``Lyman drops'' and bluer
UV continuum, thus getting closer to the locus of general field
galaxies.  As discussed earlier, such observations are useful for
exploring the dependence of the redshift distribution function of
the confirmed LBGs on the details of the color selection, as well
as for measuring contamination by low--redshift interlopers.  In
what follows we refer to these more loosely defined \wb-- and \wv--band
dropouts simply as ``fillers''. Only three such
spectroscopically-identified fillers are considered below, two
\wb--dropout fillers (GDS~J033234.40-274124.3 at $z=3.418$, QF=B
and GDS~J033251.81-275236.5 at $z=3.468$, QF=A) and one \wv--dropout
filler (GDS~J033239.82-275258.1 at $z=5.543$, QF=C).  We will report
more extensively on these tests in following papers, which will
also include spectroscopic observations of GOODS galaxies obtained
with different instrumental configurations.

\section{FORS2 Spectroscopic Observations}

The details of the observations, including journals of the observing
runs, data reduction, the extractions of the spectra have been
reported in Vanzella et al.\ (2005, 2006, 2008), and we refer the
reader to those papers.  We recall that the wavelength coverage was
typically 5700\AA-10000\AA~with a spectral resolution of 
$R= \lambda/ \Delta\lambda = 660$, corresponding to 13\AA~at 8600\AA.  No order
separation filter was used.

In the vast majority of cases, the redshift has been calculated
through the identification of prominent features of LBG spectra,
e.g.,  \Lya\ either in emission or absorption, and Si\,{\sc ii}
1260\AA, O\,{\sc i}+Si\,{\sc ii} 1302\AA~(a blend at the spectral
resolution of our instrumental setup), C\,{\sc ii} 1335\AA, Si\,{\sc
iv} 1394,1403\AA, Si\,{\sc ii} 1527\AA, C\,{\sc iv} 1548, 1551\AA~in
absorption.

Redshift determinations have been made based on visual identification
of spectral features as well as by cross--correlating the observed
spectra against high--fidelity LBG templates of differing spectral
types using the {\em rvsao} package in the IRAF environment.  In
particular, we used emission and absorption line LBG templates from
\citet{shapley03} as well as the lensed absorption line LBG cB58
(Pettini et al. \ 2000).  Each two-dimensional spectrum has been visually inspected,
including consideration of its slit orientation on the sky.  In
many cases where no continuum has been detected, we derive a redshift
measurement from \Lya\ emission.

We have co--added all repeated spectra to improve the final S/N.
The typical exposure time for each mask was about 14,400 seconds and
for co-added sources, total exposure times range from 20,000 to
80,000~sec \citep[e.g., see][]{vanz08}.

We have assigned each measured redshift a quality flag (QF), with
values of either A (unambiguous identification), B (likely
identification; e.g., based on only one line or a continuum break),
or C (uncertain identification).  The presence of \Lya\ emission
in the second order spectrum (at $>$ 10,000\AA), has also been used
on occasion, especially for faint sources with low QFs based on
absorption features.  For example, one \wb--band dropout source
(GDS~J033221.05-274820.5) shows an apparently featureless continuum
with a second-order emission line at $\sim$ 10400\AA, 
implying $z \sim$ 3.3.  
Indeed, recently the VIMOS spectroscopic observations have
confirmed this galaxy to be at $z = 3.385$ (Popesso et al.\ 2008).

We have assigned a redshift to 118 galaxies of the initial list of
202 targets, or 58.4$\%$ of the input list; this relatively low
success rate is, in large part, due to two factors: (i) the target
list includes a relatively large fraction of faint sources --- 65
or 32.2\% of the sample have \wz$>26$; and (ii) the difficulty in
deriving redshifts for galaxies at $z < 3.6$ with our instrumental
configuration.  In the latter case, depending on the slit position,
\Lya\ and the UV absorption features are often blueward of the
spectral range available.

Of the 118 spectroscopically identified sources, 106 have redshifts
in the expected range for their adopted color selection.  Note that
some of these redshifts have already been published in \citet{vanz05,
vanz06}.  Of the sources outside the expected redshift range, one source is a low--redshift galaxy from the \wb--band
dropout sample, one is a low--redshift galaxy from the \wv--band
dropout sample, and 10 are Galactic stars (1, 3 and 6 from the \wb--,
\wv--, and \wi--dropout samples, respectively).  We note that one
faint star, GDS~J033238.80-274953.7 (\wz\ = 25.16), that we
spectroscopically classify with QF=C, has been confirmed Galactic in nature due to the
detection of its proper motion (M. Stiavelli, private communication).
Excluding the Galactic stars and the two low--redshift interlopers,
the final list of spectroscopically identified LBGs includes 46
\wb--band dropouts, 32 \wv--band dropouts, and 28 \wi--band dropouts
(reported in Tables~\ref{tab:Bdrops}, \ref{tab:Vdrops} and
\ref{tab:Idrops}).

As mentioned above, we have assigned redshifts to three high--redshift
filler targets found to be in the same redshift range as the primary
LBG sample.  This brings the total number of high-redshift ($z >
3.1$) spectroscopic identifications to 109, of which 32 have QF=C.
Of these 109 galaxies, 70 have redshift $z>4$ (24 with QF=C); 37
have redshift $z>5$ (13 have QF=C); and 32 have redshift $5.5<z<6.5$
(11 with QF=C; see Table~\ref{tab:high-z} for a summary).

Finally, we also found five serendipitously--identified, high--redshift
galaxies.  These fell, as second or third sources, on slitlets
assigned to other primary targets.  For four of them the redshift
identification relies upon a \Lya\ emission line; only in one case
does the redshift rely upon absorption features. These five
serendipitous sources are, in right ascension order:

\begin{itemize}

\item{GDS~J033218.27-274712.0, at $z=4.783$ (QF=C), marked in
Figure~\ref{Vdrop} with a red pentagon, is close to the \wv\--band
dropout selection window (\wv-\wi$>$1.901, \wi-\wz=0.501).}

\item{GDS~J033219.41-274728.4, at $z=3.250$ (QF=C). This galaxy shows
a flat continuum and an absorption doublet interpreted as C\,{\sc
iv} 1548, 1551\AA.}

\item{GDS~03322.89-274521.0, at $z=5.128$ (QF=C).  This source,
clearly visible in the \wi\ band (from which the coordinates were
measured) is not detected in the \wb, \wv, or \wz\ bands.  We assume
that the emission line is \Lya, though lacking firm constraints on
the continuum SED, we can not rule out another interpretation such
as [O\,{\sc ii}]3727 at redshift $z=0.999$.  This source is not
used in the following analysis.}

\item{GDS~J033228.94-274128.2, at $z=4.882$ (QF=B), is discussed
in \citet[][see their Fig.~13]{vanz05}.  The source is not present
in the ACS catalogs because of blending with a bright galaxy.}

\item{GDS~J033243.16-275034.6, at $z=4.838$ (QF=C), is discussed
in \citet[][see their Fig.~2, top panel]{vanz06}.  This source is
not present in the ACS catalogs because of blending with a bright
star.}

\end{itemize}

Figures~\ref{1D_part1} and \ref{1D_part2} show the one-dimensional
spectra for all confirmed LBGs, separated depending on whether \Lya\
is in emission or absorption.  Figure~\ref{2D_part1} shows the
two-dimensional spectra of confirmed LBGs at $z > 5$.
Table~\ref{tab:high-z} summarizes the characteristics of each dropout
sample compared with those expected from the Monte Carlo simulations
of the redshift selection described in \S2.2, while Figure~\ref{fig:zdistr}
shows the observed redshift distribution of each dropout category.
Note the (small) overlap between the redshift distribution functions
for \wb-- and \wv--band dropouts at $z \sim 4.5$ and between \wv--
and \wi--band dropouts at $z \sim 5.5$.

\section{Efficiency of the Photometric Selections}

The effectiveness of the LBG color selection has been verified at
$z \sim 3$ by means of an extensive program of spectroscopic
confirmations of over a thousand $U$--band dropouts
\citep{steidel03}.  At higher redshift, the spectroscopic samples
of LBGs collected by various groups \citep[e.g.,][]{steidel99,
vanz06, vanz08, popesso08, yos06, ando07} are rather small and
estimates of successful identification
rates for a given set of color criteria remain correspondingly
uncertain.  Details of the filters and color criteria used by various
surveys can result in different relative proportions of
successfully-identified LBGs (i.e., in the targeted redshift range),
interlopers (i.e., outside of the targeted redshift range), as well
as other types of unwanted sources (e.g., AGN in the targeted
redshift range). The GOODS data set is being widely used for a
variety of studies of the properties of galaxies at high redshifts
(e.g., Bouwens et al.\ 2007), and many of these studies use samples
of photometrically--selected, high--redshift galaxies from the GOODS
ACS data without spectroscopic verification of the effective
composition of the samples. Even under the most optimistic assumption
that the samples include negligible fractions of interlopers and
unwanted sources, fundamental quantities such as the shape of the
redshift distribution, which is important for the measures of the
spatial clustering and luminosity function, remain largely unknown.

The spectroscopic sample obtained with FORS2 discussed here is our
initial effort to characterize the effectiveness of the GOODS LBG
color selection criteria in selecting star--forming galaxies at
high redshifts; the numbers cited below are summarized in
Table~\ref{tab:high-z}.

Of the 85 \wb--band dropout candidates selected for spectroscopic
observations, we have secure redshifts for 48 sources down to \wz\
$= 25.5$ (56\%).  Of the 48 identifications, 46 have redshifts in
the expected range for \wb--band dropouts, and only two are foreground
objects. One is a Galactic star (QF=B, \wz=23.43, SExtractor
stellarity index S/G=0.99) and the other is a galaxy at $z=1.541$
identified from [O\,{\sc ii}] 3727 emission (QF=B, \wz=25.49)\footnote{
This galaxy is well detected in the \wv, \wi\ and \wz\ bands 
(with signal to noise ratios of 13.9, 14.1 and 26.1,
respectively). This fact excludes the \Lya\ possibility at redshift $\sim$ 6.8.
Other possible emission lines are [O\,{\sc iii}]5007 and/or H$\beta$ at z $\sim$ 0.9, 
however in this case the [O\,{\sc ii}] 3727 should be detected at 7049\AA, 
region free from skylines. If we assume H$\alpha$ line, we should expect to observe 
at least H$\beta$ at 7014\AA. Therefore the most probable interpretation is [O\,{\sc ii}] 3727 
at z=1.541.}.
We have classified 8 of the 48 identifications as having QF=C.  Assuming
that all the identifications with QF=C are correct, the efficiency
of the \wb--band dropout selection
is 46/48=96$\%$.  Omitting the two foreground sources,
the mean and rms of the \wb--band dropout redshift distribution are
$\langle z \rangle=3.765$ and $\sigma_z$=0.328, respectively, fully
consistent with the prediction from Monte Carlo simulations (see
\S2.2 and Table~\ref{tab:high-z}).

\begin{figure}
 \plotone{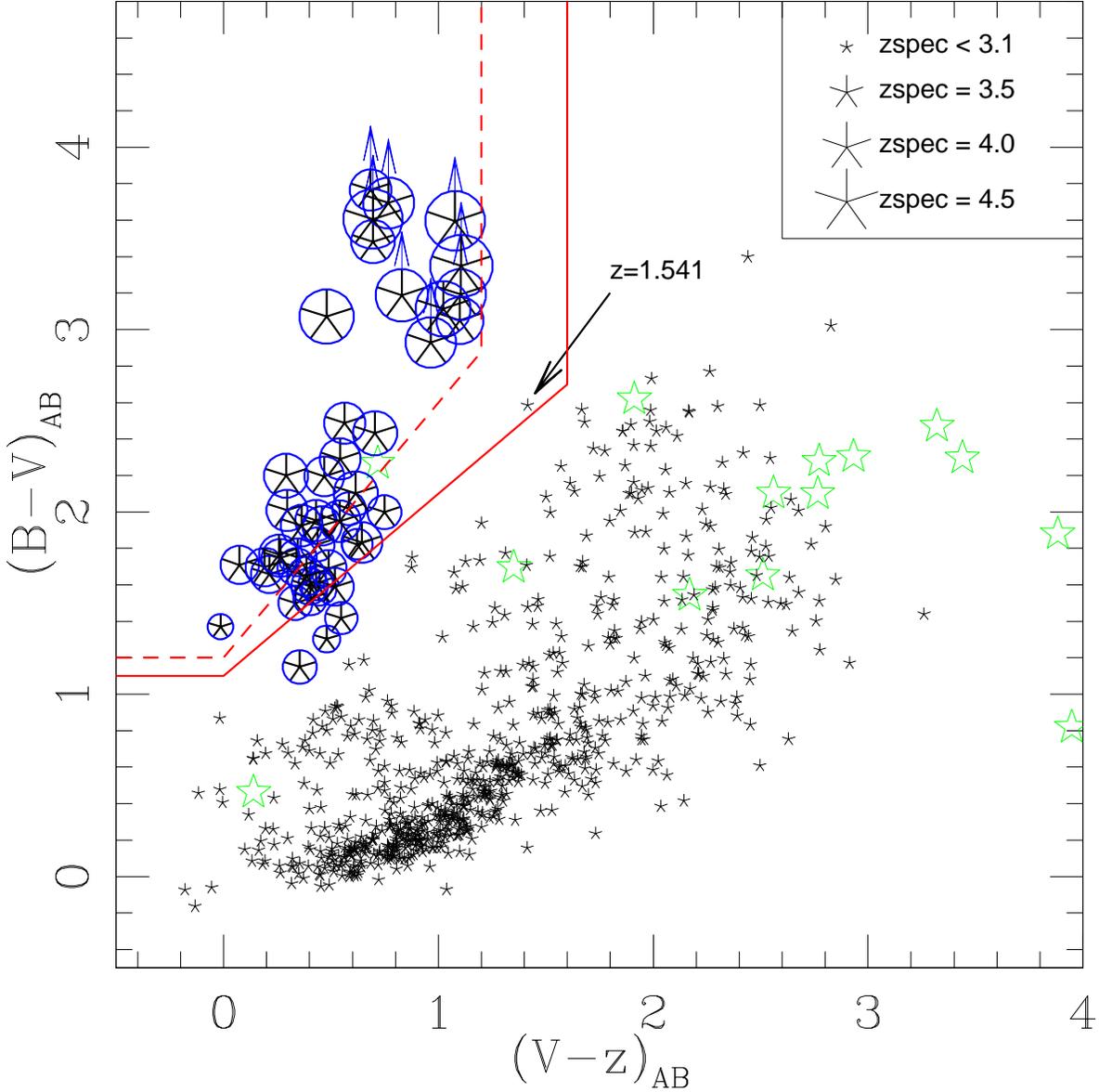} %Bdrop_new.ps
\caption{Color-color diagram for the selection of \wb--band dropout galaxies, the
solid line outline the region of the selection. The black ``skeletal'' symbols with fixed size 
are all sources in the FORS2 sample in the redshift range $0<z<3.1$, those with 
redshift in the range $3.1<z<4.4$ are plotted varying the symbol size
accordingly with the spectroscopic redshift value. Stars have
been marked with ``star'' green symbols.
Galaxies confirmed in the redshift interval $3.1<z<4.4$ have been marked with
open circles. The arrows mark one sigma lower limit of the colors.  The one low-z galaxy
 identified at z=1.541 has been marked with an arrow (see text for details).\label{Bdrop}}
\end{figure}

Figure~\ref{Bdrop} shows the (\wb-\wv) vs. (\wv-\wz) color--color
diagram for the entire FORS2 spectroscopic sample.  The region of
\wb--band dropout sources is marked with a solid line and the size
of the symbols scale linearly with redshift for sources at $z >
3.1$; at redshift lower than 3.1, the size of the symbol is fixed.
The majority of the galaxies with $z>3$ lie in the \wb--band dropout
region.  It is evident from Figure~\ref{Bdrop} that the lower tail
of the redshift distribution is located in the lower part of the
selection region:  the eight sources with $3.1<z<3.5$ have a mean
(\wb-\wv)=1.69$\pm$0.23 and (\wv-\wz)=0.44$\pm$0.21.  In this
redshift range, the selection criteria are more uncertain and depend
on the intrinsic properties of the sources.   Photometric errors
may also scatter sources across the boundary of the selection region.

\begin{figure}
 \plotone{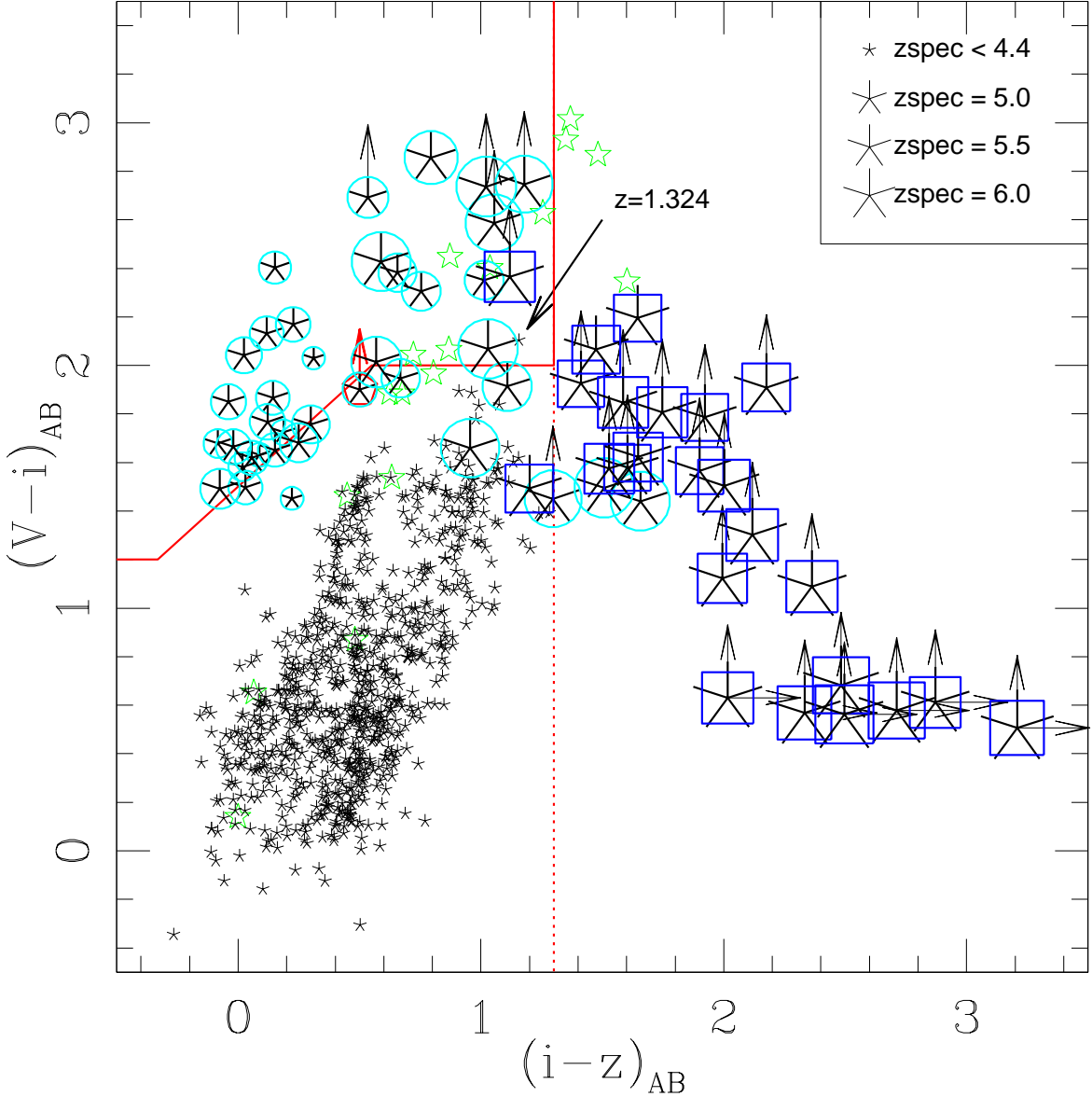} %Vdrop_new.ps
 \caption{Color-color diagram for the selection of \wv--band dropout galaxies and
 \wi--band dropout ones, the solid line outline the region of the \wv--band dropout
 selection, while the vertical dotted line ouline the \wi--band dropout
 region (\wi\-\wz\ $>$ 1.3). The black ``skeletal'' symbols with fixed size 
 are all sources in the FORS2 sample in the redshift range $0<z<4.4$, those with 
 redshift in the range $4.4<z<6.5$ are plotted varying the symbol size
 accordingly with the spectroscopic redshift value.
 Galaxies confirmed in the redshift interval $4.4<z<5.6$ have been marked with
 open circles and those with z$>$5.6 have been marked with open squares. Stars have
 been plotted with ``star'' green symbols.
 The open pentagon marks a serendipitously discovered galaxy (z=4.783 QF=C, see text).
 The arrows mark one sigma lower limit of the colors. The one low-z galaxy
 identified at z=1.324 has been marked with an arrow (see text for details).
\label{Vdrop}}
\end{figure}

\begin{figure}
 \plotone{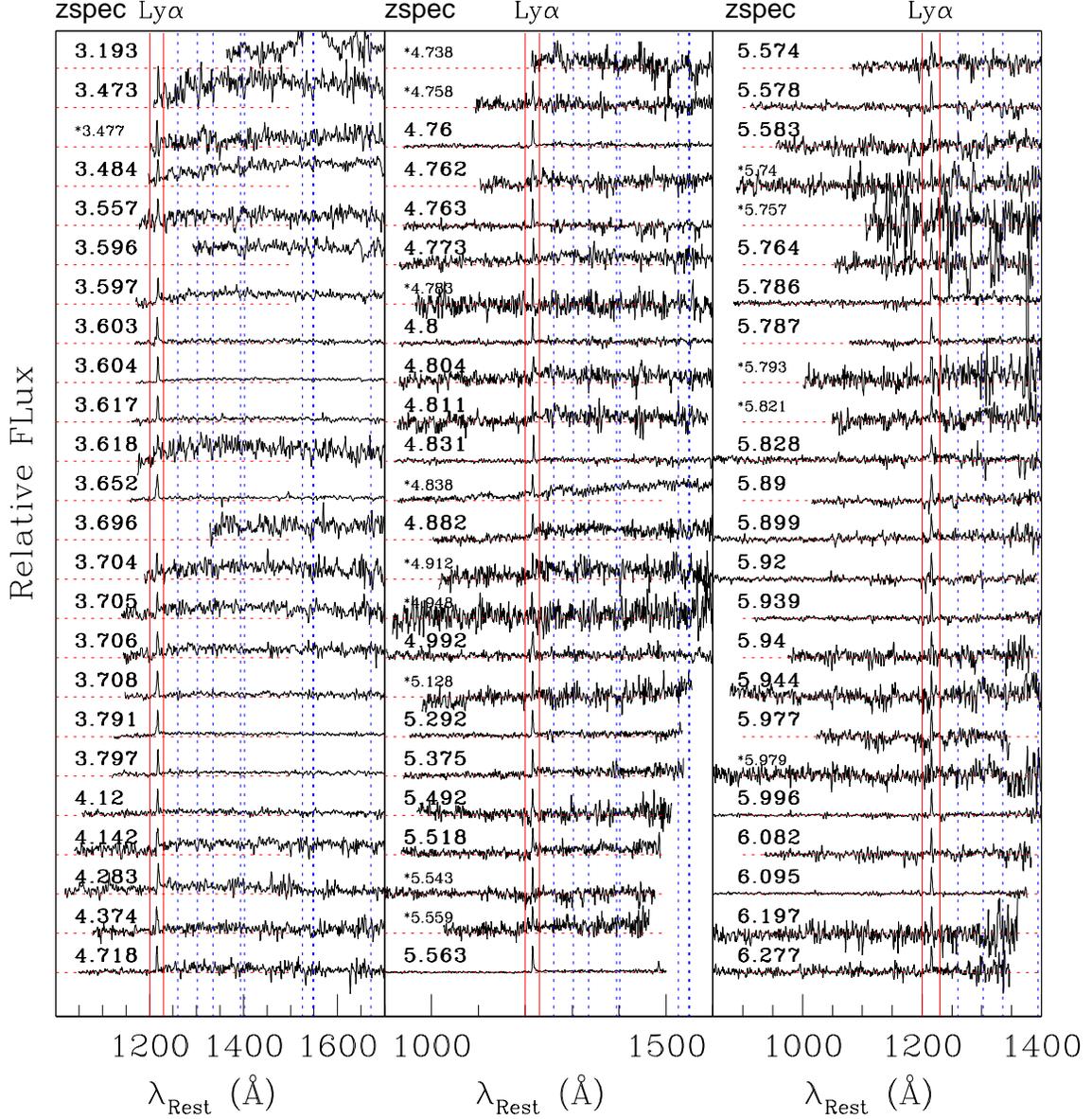} %SPECT_1Dem.ps} 
 \caption{One dimensional FORS2 rest frame spectra of all emission line galaxies of the present sample.
The redshift is indicated in the left side and the \Lya\ emission line is enclosed between the two vertical lines.
Quality C (see text) redshifts are marked with the $*$ symbol. Dotted vertical lines from left to right mark 
Si\,{\sc ii} 1260\AA, O\,{\sc i}+Si\,{\sc ii} 1302\AA, C\,{\sc ii} 1335\AA, Si\,{\sc iv} 1394,1403\AA,
Si\,{\sc ii} 1527\AA, C\,{\sc iv} 1548, 1551\AA~ and Al\,{\sc ii} 1670\AA~in absorption, respectively.\label{1D_part1}}
\end{figure}

\begin{figure}
 \plotone{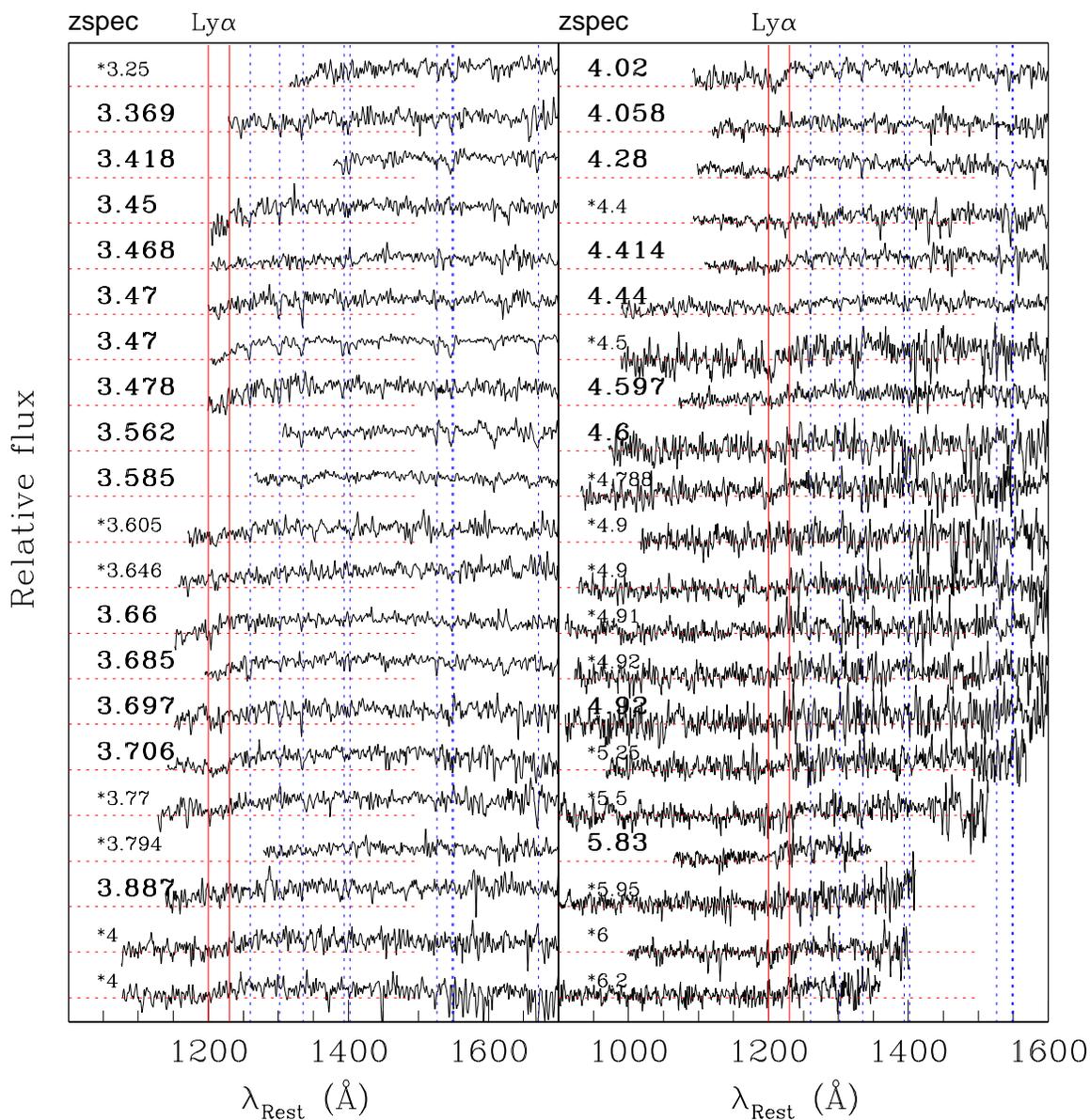} %SPECT_1Dabs.ps} %
 \caption{As in Figure~\ref{1D_part1} the one dimensional FORS2 rest frame spectra of all absorption line galaxies are shown.
 Beyond redshift $\sim$ 4.5 the spectra appear more noisy and the line identification is visually instable, 
 the cross correlation technique is particularly useful in these cases. The position of the \Lya\ line or the starting
 decrement by the IGM, is shown with solid vertical lines. Dotted vertical lines as in Figure~\ref{1D_part1}. \label{1D_part2}}
\end{figure}

\begin{figure}
 \epsscale{0.85}
 \plotone{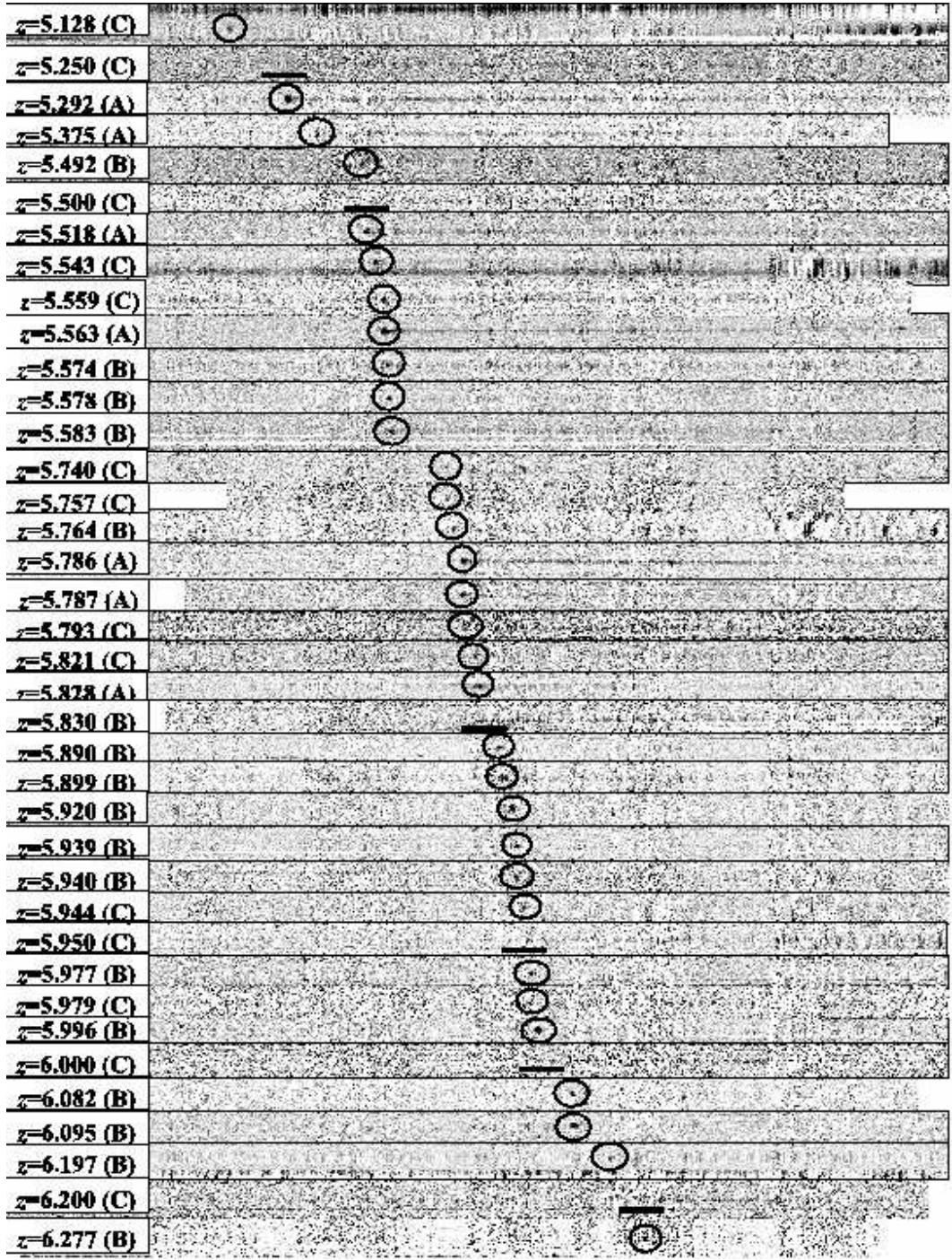}  %2D_SPEC_LBGs_gt5_NEW_2008.eps}
 \caption{Two dimensional FORS2 spectra of galaxies at redshift greater 5. 
The redshift with its quality flag is indicated in the left
side. The \Lya\ emission line is marked with a circle where
present, otherwise a segment underline the possible
continuum-break.\label{2D_part1}}
\end{figure}

\begin{figure}
 \plotone{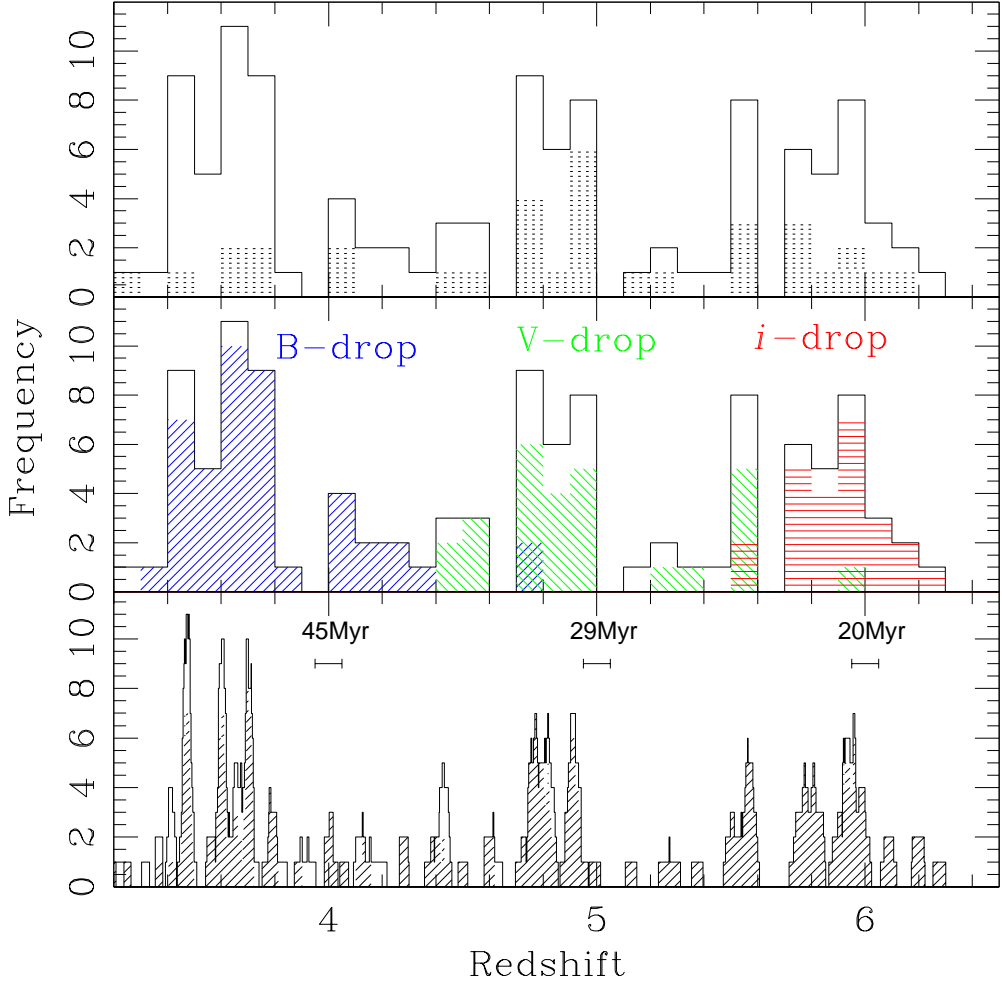} %hist_highz.ps}
\caption{Redshift distribution of the LBGs spectroscopically confirmed in the
GOODS-S field. {\bf Upper panel:} the redshift distribution of all sources at
redshift beyond 3 discovered during the FORS2 campign is shown. The dotted area
represents the sources with lower spectral quality (QF=C). 
{\bf Middle panel:} the redshift distribution (continuum line) of the FORS2
sample with the highlighted the categories \wb\ , \wv\ and \wi--band dropouts 
(blue hatched ``/'' lines, green hatched ``$\backslash$'' lines and red 
horizontal lines, respectively) is shown. 
{\bf Bottom panel:} the redshift distribution 
has been calculated counting the number of sources in a redshift bin of 0.1 and moving 
it with a step of 0.003 up to redshift 6.5 (the shaded region is the FORS2 
spectroscopic sample and the continuum line histogram include the spectroscopic 
data from the literature (see text)). The three segments indicate the interval
of cosmic time for $dz=0.1$ at the mean redshift of each category.
\label{fig:zdistr}}
\end{figure}

For the sample of \wv--band dropouts, we have assigned spectroscopic
slits to 52 candidates and derived a redshift identification for
36 of them (69\%) down to \wz$= 26.7$, of which 11 have been given
QF=C. Among the confirmed redshifts, 32 are in the range expected
for \wv--band dropouts (11 with QF=C), three are Galactic stars
(all of them with \wz$\sim 23.5$ and S/G=0.99), and one,
GDS~J033220.31-274043.4, is a low--redshift interloper at $z=1.324$
(QF=B)\footnote{This galaxy is interesting in its own right. It has
colors (\wv-\wi)=2.077 and (\wi-\wz)=1.246 and falls in the upper
right portion of the selection region in Figure~\ref{Vdrop}.  A
pronounced break around 4000~\AA\ and the Ca\,{\sc ii} HK absorption
lines are evident in the spectrum, but there is no [O\,{\sc ii}]
3727 emission line identified down to a $3\sigma$ of 2$\times10^{-18}$
erg~s$^{-1}$~cm$^{-2}$~\AA$^{-1}$, suggesting that the emission is
dominated by evolved stellar populations with little or no star
formation activity.}.  Assuming that all of the QF=C identifications
are correct, the efficiency of the $z\sim 5$ LBG selection is
32/36=89$\%$; omitting the foreground sources, the mean and rms of
the redshift distribution are $\langle z \rangle=4.962$ and
$\sigma_z$=0.386, respectively.  Again, the observed redshift
distribution agrees well with the Monte Carlo predictions
(Table~\ref{tab:high-z}).

Figure~\ref{Vdrop} shows the (\wv-\wi) vs. (\wi-\wz) color--color
diagram for the entire FORS2 spectroscopic sample.  The selection
window for the \wv--band dropouts (solid lines) and \wi--band
dropouts (dotted line; \wi--\wz$>$1.3) are plotted.  Galaxies
confirmed in the redshift interval $4.4 < z < 5.6$ are marked with
open circles.  The majority of galaxies at $z>4.4$ are located
within the selection regions.

Finally, of the 65 \wi--band dropouts selected for spectroscopic
observations, we have secured redshifts for 34 down to \wz$= 27.4$
(52\%). Of these, 28 have redshifts in the range $5.5<z<6.3$, of
which 23 are based on the identification of an observed emission
line as redshifted \Lya\ (seven have QF=C) and five are based on
the identification of an observed continuum break as the onset of
the high--redshift \Lya\ forest (four have QF=C and one has QF=B).
The remaining six \wi--band dropouts are Galactic stars (four with
QF=C and two with QF=B; all with stellarity index S/G$>$0.91 and \wz$<$25.4).
Assuming that all of the QF=C identifications are correct, the efficiency
of the $z\sim 6$ LBG selection is thus 28/34=$82\%$.  The average
redshift and standard deviation of the successfully identified
\wi--band dropouts are $\langle z \rangle$=5.898 and $\sigma_z$=0.184.
While the average of the distribution is consistent with the predicted
one, we note that the standard deviation is almost a factor of two
narrower.  This may be an indication of large--scale structure
at this redshift; from ACS grism spectroscopy, \citet{mal05} note
structure at this same mean redshift in the HUDF.

We note that several sources from our \wi--dropout spectroscopic
sample were previously published, including spectroscopic observations.
One has a well-observed spectrum showing \Lya\ emission at $z=5.829$
\citep{stan04a, dick04, bunk04}. An additional two sources were
observed with the low--dispersion ACS grism and show strong spectral
breaks interpreted as due to the \Lya\ forest at z$\sim$5.9
\citep{mal05}.  These galaxies are present in our list with redshifts
$z=5.92$ and 5.95 (see Table~\ref{tab:Idrops}).  The source
GDS~J033234.55-274756.0, for which the FORS2 spectrum yielded an
inconclusive redshift determination, is presented in \cite{mal05}
at redshift $z=6.1$.

Figure~\ref{Vdrop} shows the selection diagram for the \wi\--band
dropouts (dotted line).  Galaxies confirmed in the redshift interval
$5.6 < z < 6.5$ are marked with open squares.  The majority of
galaxies at $z>5.6$ have (\wi-\wz) redder than 1.3, as per the
adopted selection criteria.  The confirmation of galaxies at redshift
beyond five are almost exclusively due to the presence of a single
emission line, identified as \Lya\ (see \S5.4 for a dedicated
discussion).

In the redshift interval $5.4 < z < 5.6$, the \wv-- and \wi--band
dropout selection criteria overlap. In this redshift range, five
spectroscopically--confirmed galaxies meet our \wv--band dropout
selection criteria, while two meet the \wi--band dropout criteria.
As discussed below, the presence of the \Lya\ emission line may
play an important role in this respect.

\begin{figure}
 \epsscale{1.0}
 \plotone{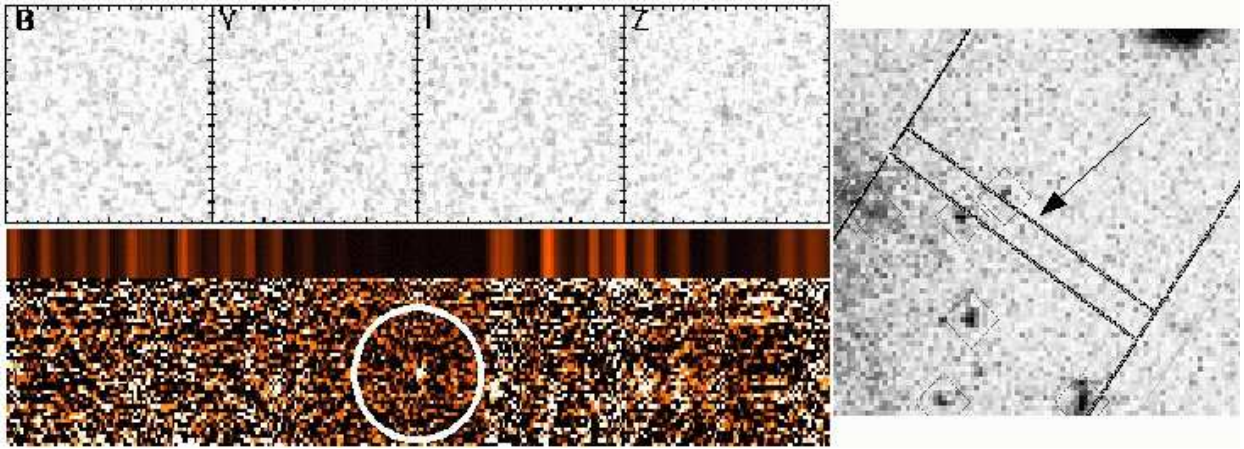} %z5p74_ALL3.ps}
 \caption{Cutouts, slit position on the sky and the two dimensional extracted spectrum 
 of the source GDS~J033233.52-275532.2 (not detected in the v2.0 ACS catalog).
 Cutouts from left to right are \wb, \wv, \wi~and \wz, respectively, with 2 arcsec side box.
 On the right part of the figure the slit position is shown, and the faint source in the center,
 indicated by the arrow (in the images north is up and east on the left). 
 On the bottom-left, the two dimensional spectrum is shown with the spot (marked with a circle) 
 in the middle of the sky window at $\sim$8200\AA~(see also Figure~\ref{LYA_SKY}) and tentatively
 interpreted as \Lya\ emission, QF=C. \label{z5p74}}
\end{figure}

Finally, as can be seen in Figures~\ref{Bdrop} and~\ref{Vdrop}, of
114 high--redshift galaxies (109 targeted and five serendipitous),
12 are outside of the primary color selection windows. Three of
them are the above mentioned ``fillers'', five are serendipitous
sources discussed in \S3, and nine are galaxies with colors close
to the \wb--, \wv-- or \wi--band dropout selection windows (the
mean ``distance'' in terms of color from the selection windows is
$\Delta C$ $\sim$ 0.04).  These galaxies were selected as \wb--,
\wv-- or \wi--band dropouts from the previous (v1.0) ACS catalog,
though in the current v2.0 catalog, they no longer meet the dropout
selection criteria (see Table~\ref{tab:99drops}).  We further note
that seven out of these nine galaxies have been classified with
QF=C.  In particular, GDS~J033233.52-275532.2, an \wi\--band dropout
at redshift $z=5.74$ (QF=C), satisfied the \wi--band dropout selection
criteria using the v1.0 catalog (in the v1.0 catalog, \wi-\wz=1.791),
but not using the v2.0 catalog; it is no longer identified at S/N$>$5
in the \wz\ band.  A visual inspection of the \wz\ image suggests
a faint source, as evident in Figure~\ref{z5p74}.  However, further
investigations will be needed to clarify this target; we do not
include this source in the following analysis.

\section{Composite Spectra}

We now describe, for each category of dropout, the general spectral
properties observed. As in the case of LBGs at $z\sim 3$, and
depending on the S/N ratio, the most prominent rest--frame UV
features observed in our samples are the HI \Lya\ line (seen in
emission, absorption, or a combination of both), low-ionization,
resonant interstellar metal lines such as Si\,{\sc ii} 1260\AA,
O\,{\sc i} + Si\,{\sc ii} 1302\AA, C\,{\sc ii} 1335\AA, Si\,{\sc
ii} 1527\AA.  Fe\,{\sc ii} 1608\AA~and Al\,{\sc ii} 1670\AA, and
high-ionization metal lines such as Si\,{\sc iv} 1394,1403\AA\ and
C\,{\sc iv} 1548, 1550\AA\ associated with P-Cygni stellar wind
features and ionized interstellar gas.  In one case, N\,{\sc iv}]
1485\AA~emission has been detected together with \Lya\ in emission
(GDS~J033218.92-275302.7; Vanzella et al. \ 2006).  As to be expected, the number of
robustly identified lines decreases drastically with apparent
magnitude over the range from \wz$\sim 24$ to $\sim 26.5$.  At the
faintest magnitudes, given our typical exposure times, continuum
flux is at the limit of measurability (S/N$\sim$1 per resolution
element) and therefore no absorption lines are reliably observed;
the only observable feature for the faintest sources is \Lya\
emission.

\subsection{\wb\--Band Dropout Composite Spectra}

Among 46 \wb--band dropouts with spectroscopic redshift at $z\approx
4$, 15 of them show \Lya\ in emission line (the ``em.'' class), 21
have redshift identified by means of absorption lines only (the
``abs.''  class) --- typically SiII 1260.4\AA, CII 1335.1\AA, SiIV
1393.8,1402.8\AA, CIV 1548.2,1550.8\AA\ --- and 10 sources show
both emission and absorption features (the ``comp.'' class). As
mentioned above, two \wb\--band dropouts have \Lya\ blueward of the
observed spectral range, but this line was visible in second order
at $\lambda >$ 10,000\AA; only absorption features were used to
derive the redshifts for these sources.  These two galaxies have
not been used to make the composite spectra.

The composite spectra, normalized at 1450\AA, for emission line
sources (``em.'' class), emission and absorption line sources
(``comp.'' class) and absorption line sources (``abs.'' class) at
$z \sim$ 3.8 are shown in the left panel of Figure~\ref{fig:stackBVdrop}.
The composite spectra include sources with QF = A, B and C.  A
continuum break blueward of \Lya, due to the intergalactic medium,
is clearly evident.  Stellar and interstellar lines are also easily
recognized.

The absorption lines clearly differ between the ``em.'', ``comp.''
and ``abs.'' classes.  Figure~\ref{fig:BdropCOMP} superposes
the composite spectra of the \wb\--band dropouts with and without
the \Lya\ emission line (``em.'' and ``abs.'').  Low-ionization
interstellar absorption lines are more pronounced in the ``abs.''
class composite spectrum; e.g., compare the OI, CII and FeII lines.
Figure~\ref{fig:BdropCOMP} also shows that the non-emitter population
has a redder spectral slope, consistent with the previous work based
solely on photometric data; e.g., \citet{pente07} find $\beta_{\rm
phot}^{\rm em.} \sim -2.0 \pm 0.11$ and $\beta_{\rm phot}^{\rm abs.}
\sim -1.7 \pm 0.13$, where $F(\lambda) \sim  \lambda^{-\beta}$.  A
similar trend has also been noted by \citet{shapley03} from their
sample of $z \sim 3$ LBGs.  In particular, \citet{shapley03} find
that the average extinction, $E(B-V)$, decreases as a function of
increasing \Lya\ emission strength.  The similar trends seen here
at $z \sim 4$ suggests that the emission line \wb--dropout LBGs
are, on average, less extincted than the absorption line \wb--dropout
LBGs \citep[c.f.,][]{pente07}.

\subsection{\wv\--Band Dropout Composite Spectra}

As reported above, 32 \wv--band dropouts are at $z\approx 5$.  As
expected for the larger distance modulus and correspondingly fainter
sample, the fraction of spectroscopically--confirmed \wv--band
dropouts with \Lya\ in emission is higher compared to the \wb--band
dropout sample; 19 sources are in the ``em.'' class and 13 show
only absorption lines or a continuum break (``abs.'' class).  The
rest--frame composite spectra of emission and absorption galaxies
with QF = A, B and C are shown in the top left panel of
Figure~\ref{fig:stackBVdrop}.  Stellar and interstellar absorption
lines, as well as the strong continuum discontinuity at \Lya, are
clearly observed in the composite spectra.  The composite ``em''-class
spectrum looks quite similar to the \wb--band dropout composite
``em'' spectrum.  For the ``abs.'' class, the composite is dominated
by low quality (QF=C) spectra and only the strong Lyman-$\alpha$
forest break is apparent.

\subsection{\wi\--Band Dropout Composite Spectrum}

The rest--frame composite spectrum of the emission line (``em.''
class) \wi--band dropouts is shown in Figure~\ref{fig:stackIdrop}.
Among the 28 \wi--band dropouts with spectroscopic redshift at
$z\approx 6$, 22 sources show \Lya\ in emission (7 with QF=C).

Given the exposure times, these galaxies are generally too faint
to measure a continuum (e.g., see Figure~\ref{1D_part1} or
Figure~\ref{2D_part1}) and only \Lya\ emission has been detected.
As shown in Figure~\ref{MAG_DISTR}, this is particularly true for
the fainter \wi--band dropouts.  Nevertheless, the composite \wi--band
dropout spectrum shows signal redward of the \Lya\ line, with
tentative detection of the Si\,{\sc ii} 1260\AA~and O\,{\sc i} +
Si\,{\sc ii} 1302\AA~absorption lines despite the sky lines at these
wavelengths being stronger and denser.  At these high redshifts ($z
\sim 5.9$), we find a very opaque IGM blueward of the redshifted
\Lya\ line.  Consistent with quasar results \citep[e.g.,][]{songaila04},
the IGM transparency is estimated to be of the order of 1$\%$.

\subsection{Single--Line Redshift Identifications}

For most of the $z \simgt 5$ LBGs in our sample, the redshift
identifications are based on a single emission line --- assumed to
be redshifted \Lya\ --- in an otherwise featureless and/or low--S/N
ratio spectrum.  A question naturally arises:  how robust are these
identifications (e.g., Stern et al.\ 2000)?  To be selected as
dropouts, the broad--band SED of these galaxies must satisfy the
color--selection criteria, which require the signature of the Lyman
limit and/or \Lya\ forest blanketing.  The most plausible candidate
for an alternate identification is [O\,{\sc ii}]3727 at $z \simgt
1.0$, though \hbeta\ at $z \simgt 0.5$, [O\,{\sc iii}]5007 at 
$z\sim 0.5$ and \halpha\ at $z\sim 0.1$ are also possible.  Such
possibilities, however, will generally be inconsistent with the
broad--band colors of the galaxies, since low--redshift solutions
would be star--forming galaxies with relatively blue continua.  This
is illustrated in Figure~\ref{EW_OII_LYA}, which plots the observed
equivalent width versus the (\wi-\wz) color for [O\,{\sc ii}]--emitting
galaxies at redshift $1<z<1.5$ and \Lya--emitting LBGs at redshift
$z>5$ (only galaxies with QF=A are plotted).  Color and equivalent
width do an effective job at separatating the low--redshift,
star--forming galaxies from their high--redshift counterparts,
particularly for the \wi--band dropouts.  A few \wv--band dropouts
do overlap with the low--redshift galaxies, but are easily separated
using (\wv-\wi) color, which is better suited for galaxies at $z
\sim 5$.

\begin{figure}
 \epsscale{1.1}
 \plottwo{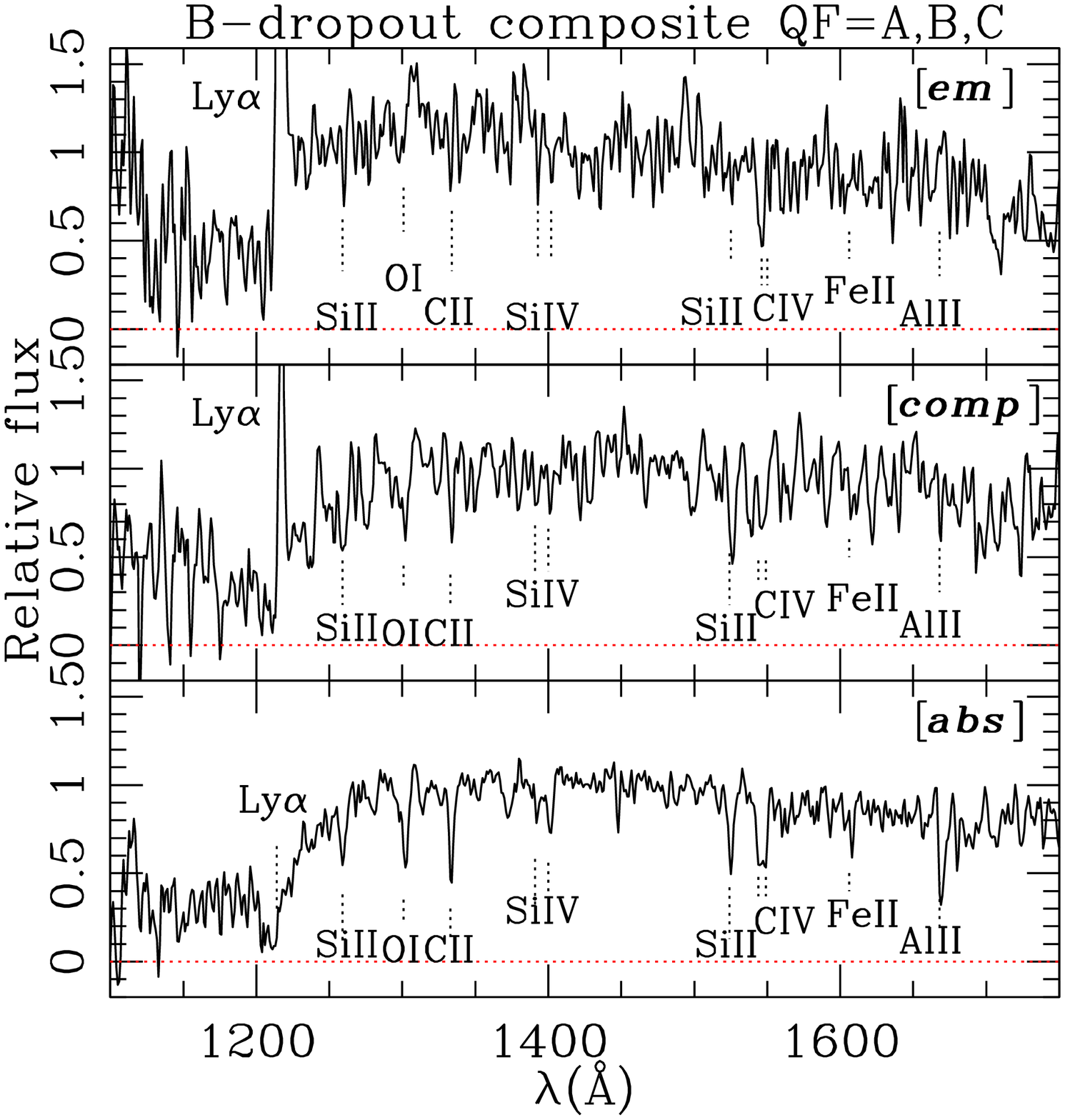}{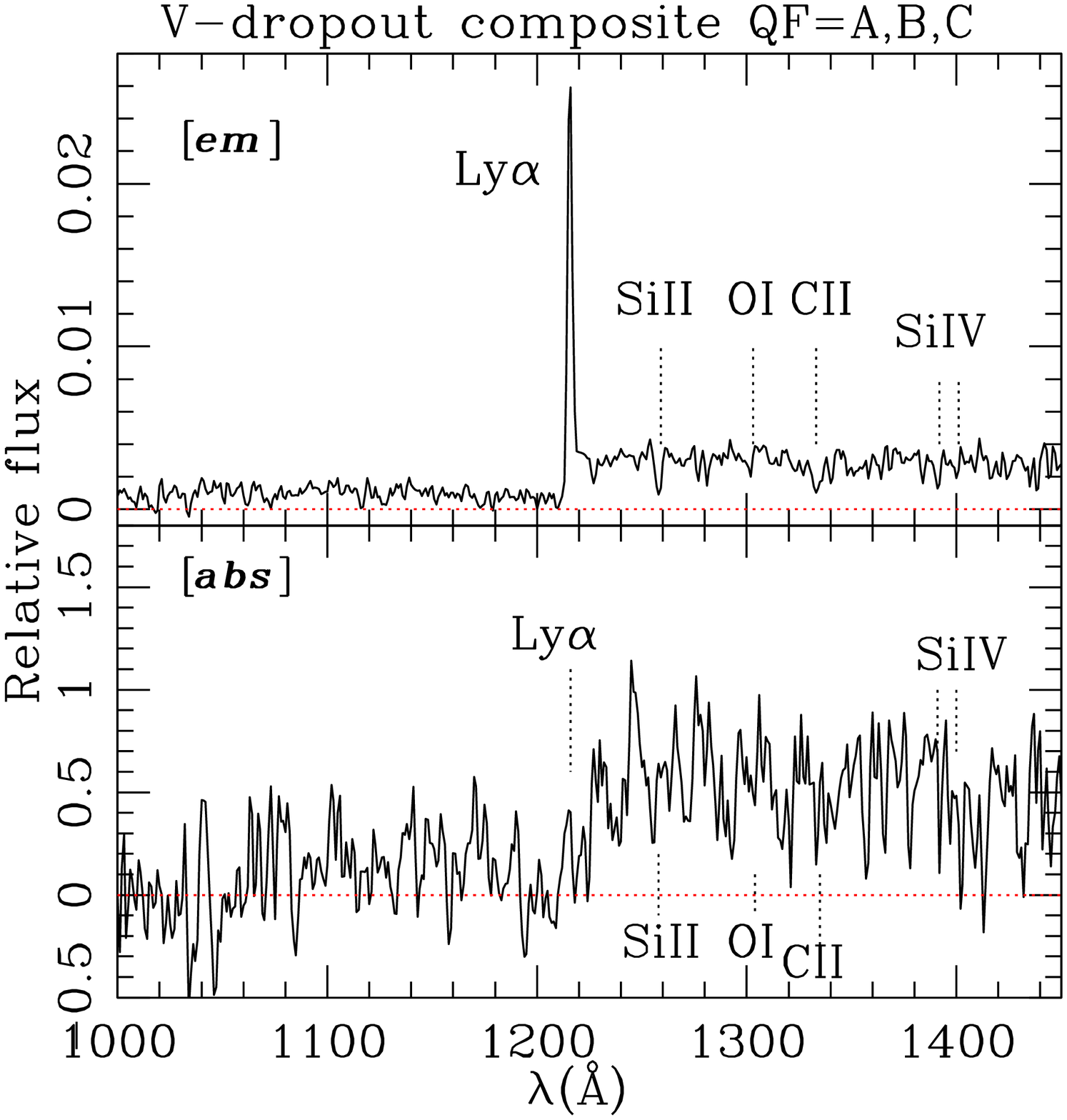} %STACK_BdropABC.ps}{STACK_VdropABC.ps}
 \caption{{\bf Left panel:} The composite spectrum of 
 \wb\--band dropout galaxies with the \Lya\ emission line (TOP), 
 \Lya\ emission and absorption features (MIDDLE) and only absorption features (BOTTOM) 
 is shown, respectively. The spectroscopic features are well recognized 
 (for a detailed comparison between emitters (TOP) and absorbers (BOTTOM)
 see Figure~\ref{fig:BdropCOMP}). 
{\bf Right panel:} The same for the \wv\--band dropout sources has been done for emitters (TOP) 
and absorbers (BOTTOM). In case of emitters, the abosrption features are
also clearly detected. In case of absorbers, given the low quality (QF=C) spectra and
the small sample, only the Lyman-$\alpha$ forest break is apparent.
\label{fig:stackBVdrop}}
\end{figure}

\begin{figure}
 \epsscale{1.0}
 \plotone{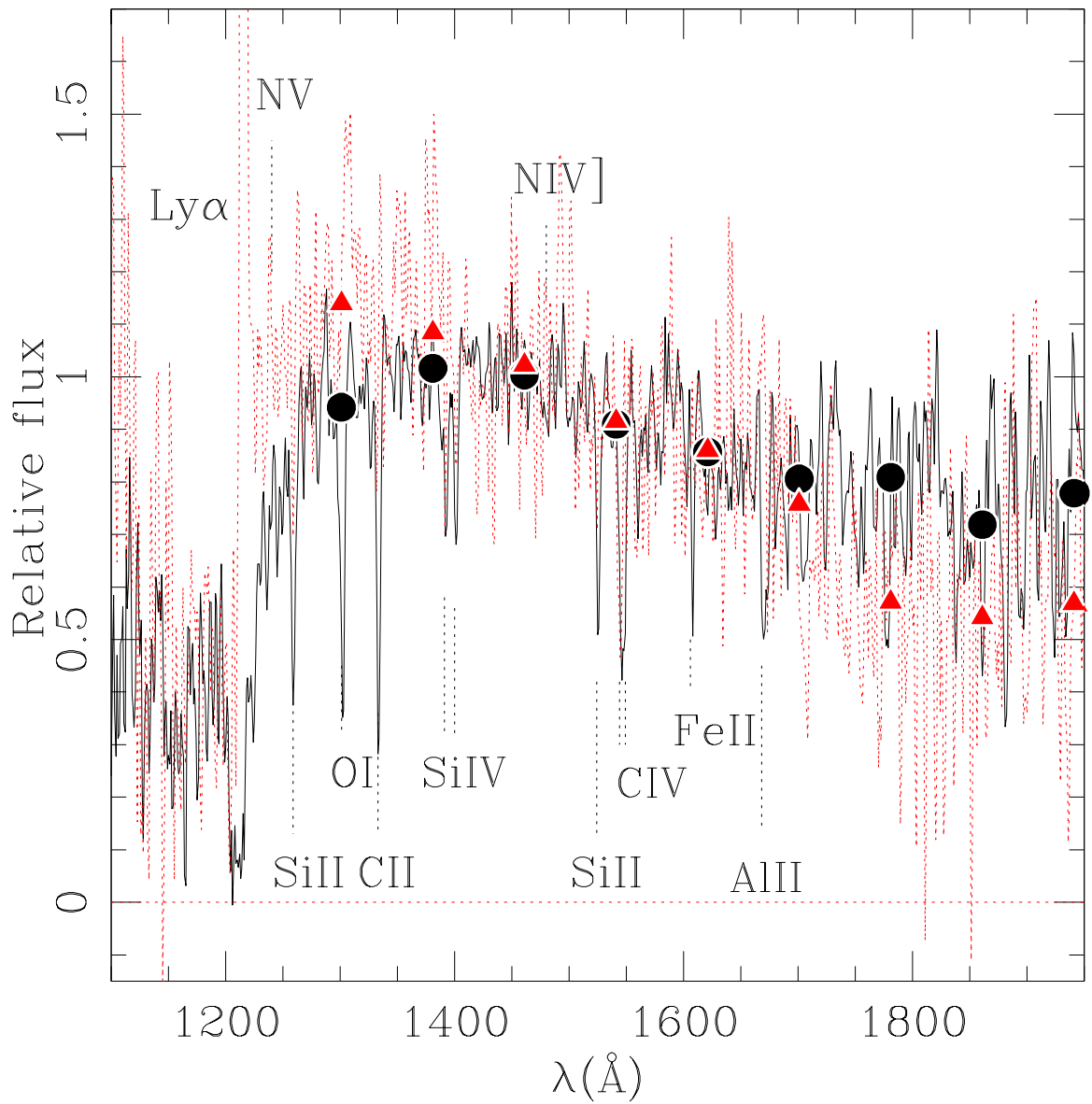} %BdropOVER.ps}
 \caption{Comparison between composite spectra normalized at 1450\AA~of the
\wb\--band dropout galaxies with and without the \Lya\ emission line (emitters and
absorbers). The circles are the median values calculated in bins of 100\AA~of the absorbed stacked
spectrum, while the triangles are those of the emission stacked spectrum. The bluer
spectral slope of the ``emitter'' population is evident and in general
the absorption lines of the emission stacked spectrum are weaker than the
absorbed spectrum.
\label{fig:BdropCOMP}}
\end{figure}

\begin{figure}
 \epsscale{1.}
 \plotone{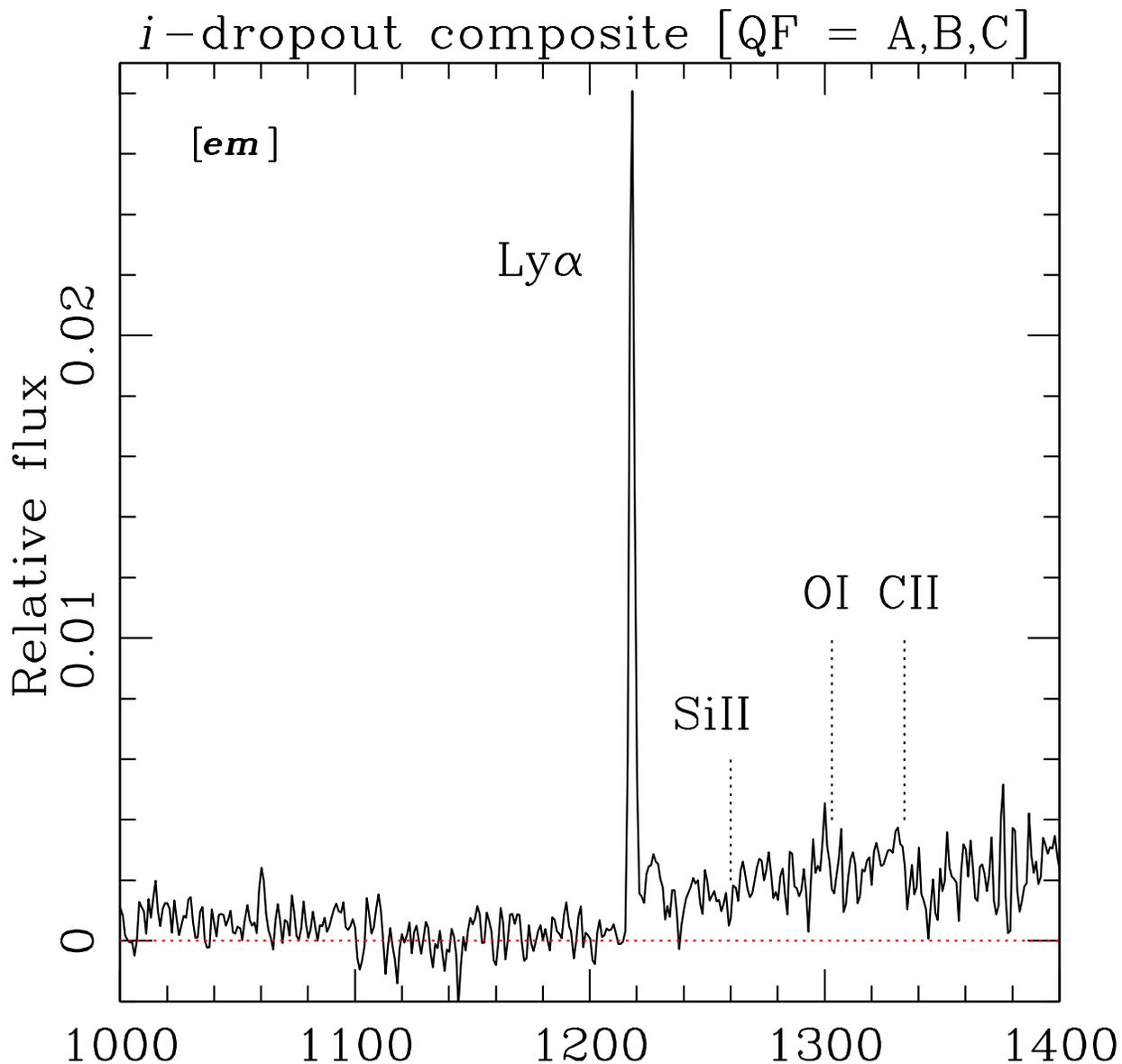} %STACK_Idrop.ps}
 \caption{Composite spectrum of \wi\--band dropout emission line galaxies.
 A faint signal reward the \Lya\ line is clear, and there is a tentative detection
 of absorption lines, whose expected position is probably disturbed by the sky lines 
 residuals (especially at $\lambda$ beyond 1340\AA).
 The shape of the spectrum shows the attenuation of the IGM blueward the \Lya\ line.
\label{fig:stackIdrop}}
\end{figure}

\begin{figure}
 \plotone{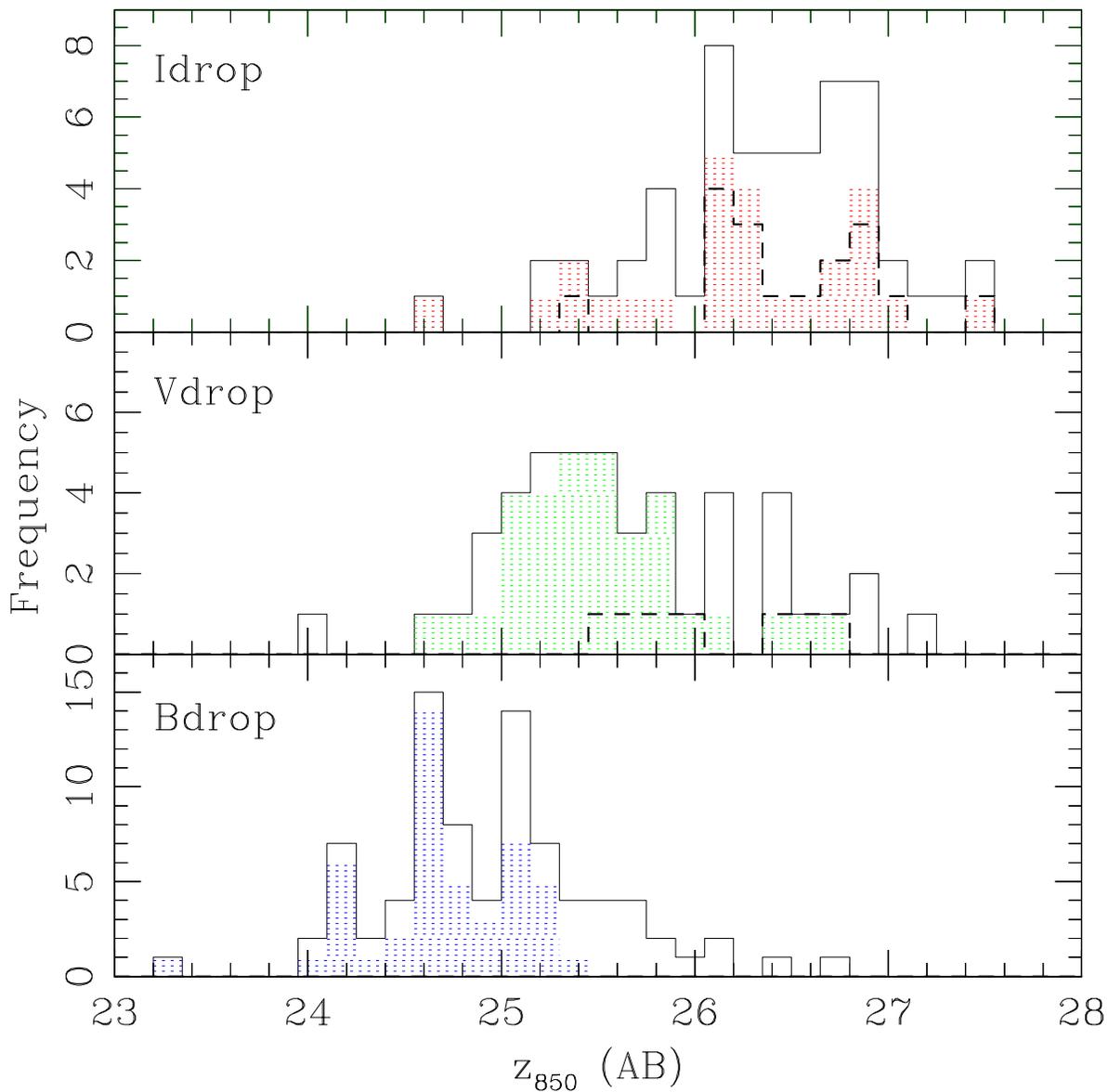} %MAGs_DISTRB.ps}
 \caption{\wz--magnitude distriburion of the three \wb\ ,\wv\ and \wi--band dropout samples. Solid line histograms 
 show the magnitude of all targets observed, dotted regions show the sources with a redshift measure and dashed histogram outline the 
 sources for which a single emissione line (without continuum) has been observed and used in 
 the redshift measurement (\Lya). 
 It is evident the single line detection for the fainter galaxies (\wz\ magnitude beyond $\sim$ 26).
 For the \wb--band dropout sample all galaxies show the continuum. \label{MAG_DISTR}}
\end{figure}

\begin{figure}
 \epsscale{1.}
 \plotone{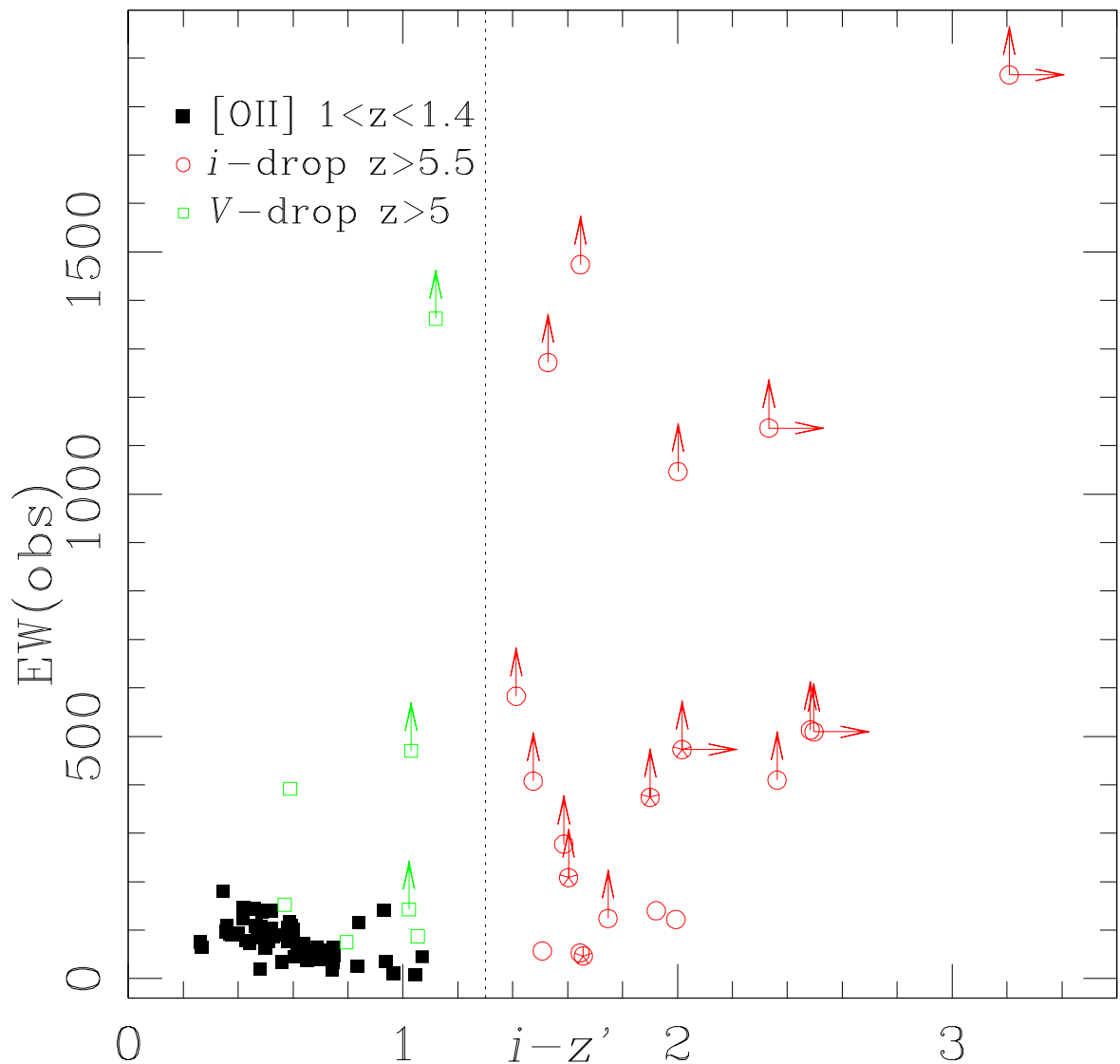} %EW_vs_iz.ps}
 \caption{Comparison of the observed equivalent widths of \Lya\ and [O\,{\sc ii}] 3727 lines for the samples of galaxies
 at redshift $>$ 5 and redshift $\sim$ 1-1.4, respectively. This is a further indication of the high redshift
 nature of the single line detected for dropouts galaxies. Three out of four \wv--band dropouts 
 relatively close to the zone of [O\,{\sc ii}] 3727 galaxies have been confirmed with QF=A, both 
 the \Lya\ line and the continuum are evident in the spectra. One is a QF=B and the equivalent widths 
 of \Lya\ is a lower limit (see text for a detailed discussion). \label{EW_OII_LYA}}
\end{figure}

The composite spectra of emission line \wv--band dropouts (Figure
\ref{fig:stackBVdrop}, right) and \wi--band dropouts (Figure
\ref{fig:stackIdrop}) also provides evidence that most of the
single--line \Lya\ identifications are correct. The $z\sim 5$
composite spectrum, which is a lower--S/N version but otherwise
virtually identical to the $z\sim 4$ composite spectrum (Figure
\ref{fig:stackBVdrop}, left), shows a number of absorption features
that would not be observed due to dilution if most of the identifications
were wrong. The $z\sim 6$ composite spectrum is similar, except
that the lower S/N ratio results in a lower--S/N detection, or no
detection at all, of absorption features. The continuum discontinuity
across the \Lya, however, is clearly detected with a jump larger
than one order of magnitude in the continuum flux density (in fact,
the continuum blueward of the \Lya\ line is consistent with being
zero).  This is larger than other continuum discontinuties observed
in distant galaxies (c.f., Spinrad et al.\ 1998; Stern et al.\
2000).

Finally, another discriminant between high--redshift and low--redshift
single emission-line sources is provided by the line profile:
high--redshift \Lya\ lines are asymmetric due to intervening H{\sc
i} absorption, while other lines will generally be symmetric.
However, the low S/N, low spectral resolution ($R$$\sim$660) reported
here makes the detection of a clear asymmetry challenging in most
individual spectra.  In a few cases, however, an asymmetric profile
has been detected in the brigher \Lya--emitting LBGs reported here.

\section{Outflows at $z\sim$ 4 and 5}

Evidence of powerful winds in LBGs at $z\sim3$ (Shapley et al.\
2003) and in galaxies at $z\sim 2$ selected from UV colors (Shapley
et al.\ 2005) has been inferred from the systematic redshift of the
\Lya\ emission line and the blueshift of interstellar absorption
lines with respect to the systemic redshift of the galaxies, as
traced by rest-frame optical nebular lines.  In this scenario the
redshifted \Lya\ emission line forms in the receding part of a
generally bipolar flow of gas, while the blueshifted interstellar
lines originate in the part along the line of sight moving toward
the observer.

\subsection{Outflows in \wb--Band Dropouts ($z \sim 4$)}

It is of interest to see if LBGs at $z\sim 4$ also show the same
phenomenon, and compare its magnitude to that of the lower-redshift
galaxies, looking for evolutionary effects.  Obtaining spectroscopic
observations of the rest--frame optical nebular emission lines is
not a trivial task.  The [O\,{\sc ii}]3727, [O\,{\sc iii}] 4959 and
5007\AA\ lines have been identified for only nine galaxies in our
sample, as a part of the AMAZE project, aimed at estimating the
mass-metallicity relation at high redshift (Maiolino et al.\ 2008).
In fact, these features become unreachable from the ground for
redshift $\simgt 3.8$ when the lines go beyond the $K$-band.   For
such sources, information about the possible presence of winds is
derived from the velocity differences between \Lya\ emission and
interstellar absorption lines.

The AMAZE project (Maiolino et al.\ 2008) has determined the redshift
of nebular lines using the integral field spectrometer SINFONI at
the VLT, adopting a spectral resolution $R$=1500 in the spectral
range $1.45-2.41 \mu$m.  For each source, the redshift derived from
[O\,{\sc iii}] 4959, 5007\AA\ and [O\,{\sc ii}]3727 agree within
$|\Delta z|$$\sim$$10^{-3}$.  We have calculated ``nebular redshifts''
for each galaxy by averaging these three lines.  The redshift of
the interstellar medium has been derived from the low-ionization
interstellar absorption lines (ISL, e.g. Si\,{\sc ii} 1260\AA,
O\,{\sc i}+Si\,{\sc ii} 1302\AA, C\,{\sc ii} 1335\AA, and Si\,{\sc
ii} 1527\AA), and the redshift of the hydrogen gas is estimated
from the \Lya\ line.

We then compare the various redshift estimates arising from the
different physical regions within the LBGs, i.e. the velocities
$V_{\rm Ly\alpha}$, $V_{\rm ISL}$ and $V_{\rm nebular}$.  We find that:

\begin{enumerate}

\item{the relative median velocity $\langle V_{\rm Ly\alpha} - V_{\rm ISL} \rangle$ 
observed between the \Lya\ emission lines and the interstellar absorption 
lines is +$370_{-116}^{+270}\, {\rm
km}\, {\rm s}^{-1}$ (derived from 16 galaxies at an average redshift
of 3.70$\pm$0.2).  The \Lya\ emission is always redshifted relative
to the interstellar lines.  Adopting the model of \cite{ver06}, the
velocity $V_{\rm exp}$ of the expanding neutral hydrogen shell is
of order of $120 - 180\, {\rm km}\, {\rm s}^{-1}$;}

\item{the relative median velocity $\langle V_{\rm Ly\alpha} -
V_{\rm nebular} \rangle$ between the \Lya\ emission line and the
nebular lines is +$161\pm80\, {\rm km}\, {\rm s}^{-1}$ (derived from
four galaxies at z $\sim$ 3.65);}

\item{the relative median velocity $\langle V_{\rm ISL} - V_{\rm
nebular} \rangle$ between the interstellar absorption lines and the
redshift of the nebular lines is $-165_{-194}^{+170}\, {\rm km}\, {\rm s}^{-1}$
(derived from nine  galaxies at z $\sim$ 3.7);}

\end{enumerate}

The galaxy GDS~J033217.22-274754.4, with its peculiar, double-peaked
\Lya\ profile is already been discussed in detail in \cite{vanz08}.

\begin{figure}
 \epsscale{1.}
 \plotone{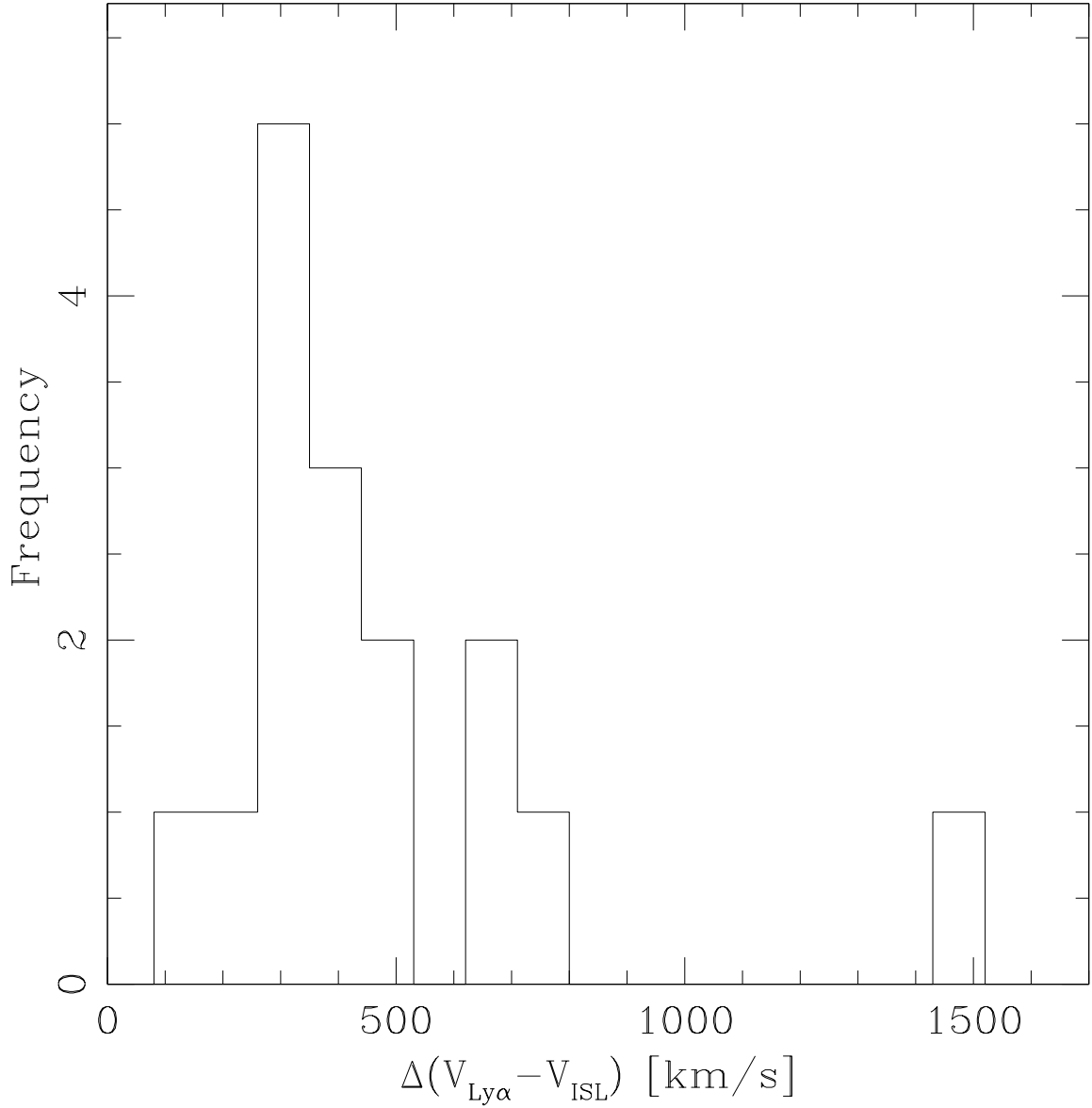} %HISTO_WIND.ps}
 \caption{Velocity differences between the the \Lya\ line and the interstellar absorption lines
($V_{Ly\alpha}-V_{ISL}$) for 16 galaxies of the \wb--band dropout sample. 
The median of the distribution is $370_{-116}^{+270}$ km/s (see text for details).
\label{histo_wind}}
\end{figure}

Figure~\ref{histo_wind} shows the histogram of $(V_{\rm Ly\alpha}-V_{\rm
ISL})$ for the 16 galaxies from the \wb\--band dropout sample for
which this measurement has been possible.  The histogram does not
include quality QF=C spectra. In all cases, the redshift of the
\Lya\ is measured by fitting a Gaussian profile to the line,\footnote{As
simulated in \cite{ver08}, the effect of the spectral resolution
on the measurement of the \Lya\ barycenter is more important for
galaxies with broad \Lya\ absorption.  In the case of emission,
like the objects reported here, this is not the case -- the lines
are narrow.  Because the lines are nearly unresolved, asymmetry has
little affect on the measured central walelength.} while the redshift
of the interstellar absorption lines is derived cross-correlating
the individual spectra with templates \citep[viz., the lensed galaxy
cB58 and the composite spectrum without \Lya\ emission from][]{shapley03},
after excluding the \Lya\ line from the analysis. The typical
redshift error is $\Delta z\sim0.001$ \citep[][derived from multiple,
independent observations]{vanz08} and translates into a final error
on the velocity difference $\Delta(V_{\rm Ly\alpha}-V_{\rm ISL})
\sim 64\, {\rm km}\, {\rm s}^{-1}$ at $z \sim 3.7$.
Figure~\ref{histo_wind} shows that the \Lya\ line is systematically
redshifted relative to the interstellar absorption lines and a few
galaxies have velocity differences in excess of $600\, {\rm km}\,
{\rm s}^{-1}$.

Though derived from relatively small samples, these numbers are
similar to LBGs at $z\sim 3$ \citep{shapley03, adel03}.  In particular,
Figure 11 of \citet{shapley03} shows that with increasing \Lya\
emission strength, the kinematic offset implied by the relative
redshifts of \Lya\ emission and low-ionization interstellar absorption
lines decreases monotonically from $\langle V_{\rm Ly\alpha} -
V_{\rm ISL} \rangle =800\, {\rm km}\, {\rm s}^{-1}$ to $\langle
V_{\rm Ly\alpha} - V_{\rm ISL}) = 480\, {\rm km}\, {\rm s}^{-1}$.
If we assume this trend remains true at $z \sim 3.7$ and consider
the mean rest--frame \Lya\ equivalent width of our sample (20 \AA),
the comparison is even more consistent with the results at $z \sim
3$.  We also note that the $\langle V_{\rm ISL} - V_{\rm nebular}
\rangle = -150\, {\rm km}\, {\rm s}^{-1}$  derived by \citet{adel03}
is similar to the value derived here at slightly higher redshift,
$-165\, {\rm km}\, {\rm s}^{-1}$.

\subsection{Outflows in \wv--Band Dropouts and at Redshifts Beyond 5}

In the case of \wv--dropouts ($z \sim 5$), the mean velocity
difference $\langle V_{\rm Ly\alpha} - V_{\rm ISL}\rangle$ is more
difficult to estimate for individual galaxies because the S/N is
generally lower, due to both the faintness of the targets and to
the UV absorption features entering a spectral region affected by
strong sky emission lines at $z \simgt 4.5$.  For this reason, we
have resorted to estimating $\langle V_{\rm Ly\alpha} - V_{\rm
ISL}\rangle$ from the composite spectrum.  All \wv--band dropouts
with \Lya\ in emission have been co-added, registering their redshift
with respect to theeir \Lya\ lines.  This procedure will lead to a
slight smoothing of the interstellar lines and thus a larger
uncertainty.  Nevertheless, an absorption signal remains clearly
detected in the composite spectrum.

The UV absorption features SiII 1260.4\AA, CII 1335.1\AA~and SiIV
1393.8,1402.8\AA~(see Figure~\ref{fig:stackBVdrop}) show an average
blueshift of $\sim -450\, {\rm km}\, {\rm s}^{-1}$  with respect
to the \Lya\ line, similar to the \wb--dropout results.  With the
aim of extending this measurement to yet higher redshift, we have
selected a subsample of eight LBGs with detected continuum at $z >
5$ from the \wv\ and \wi--band dropout samples, at an average
redshift of 5.6 and \wz\ magnitude 25.6 (three \wi--band dropouts
and five \wv--band dropouts; QF=C LBGs have not been considered).
Similar to the sample of pure \wv--band dropouts, the composite
spectrum shows a velocity offset of $(V_{\rm Ly\alpha}-V_{\rm ISL})
\sim +500\, {\rm km}\, {\rm s}^{-1}$.

In order to check if the above estimations give realistic measurements
of the offset, we have re-calculated $(\rm V_{Ly\alpha} - V_{\rm
ISL})$ from the \wb--band dropout composite spectrum.  We find
$(V_{\rm Ly\alpha} - V_{\rm ISL}) \sim +490\, {\rm km}\, {\rm
s}^{-1}$.  Though a bit higher, this value is consistent with the
number derived from individual measurements.

This analysis performed therefore supports the interpretation that
outflows at $z \sim 4$ and $5$ are present and similar to those
seen at lower redshifts ($z \sim 2-3$).

\section{\Lya\ Equivalent Width and the UV Luminosity}

For all galaxies with \Lya\ in emission, we have estimated the rest
frame equivalent width of the line. In the critical cases where
this line is the only feature detected in the spectrum, the continuum
has been estimated from the available photometry assuming a flat
spectrum with spectral index $\beta = -2.0$ ($f_\lambda \propto
\lambda^\beta$).  Depending on the redshift, the \wi\ 
($z \leq 4.65$), \wz\ ($4.65 < z \leq 5.7$) or $Js$ ($z > 5.7$) magnitudes
have been used to determine the continuum level.  In the highest--redshift
case, we use $Js$ magnitudes (the $Js$ filter has a central wavelength of 
1.24$\mu$m and width of 0.16$\mu$m, it allows an accurate photometry)
from the GOODS-MUSIC catalog \citep{grazian06}, or the NIC3 F110W band 
magnitude (Thompson et al.\ 2006) for sources in the HUDF.  
%If the \Lya\ flux derived from the spectrum is a lower limit (because the continuum is not detected),
If the magnitude is a lower limit, the resulting
equivalent width is a lower limit (indicated by an arrow in the
figures).  The absolute $M145$ magnitude has been derived from the
\wz\--band, assuming a template (drawn from SB99) of a star--forming
galaxy with spectral index $\beta \sim -2.0$.

\begin{figure}
 \epsscale{1.}
 \plotone{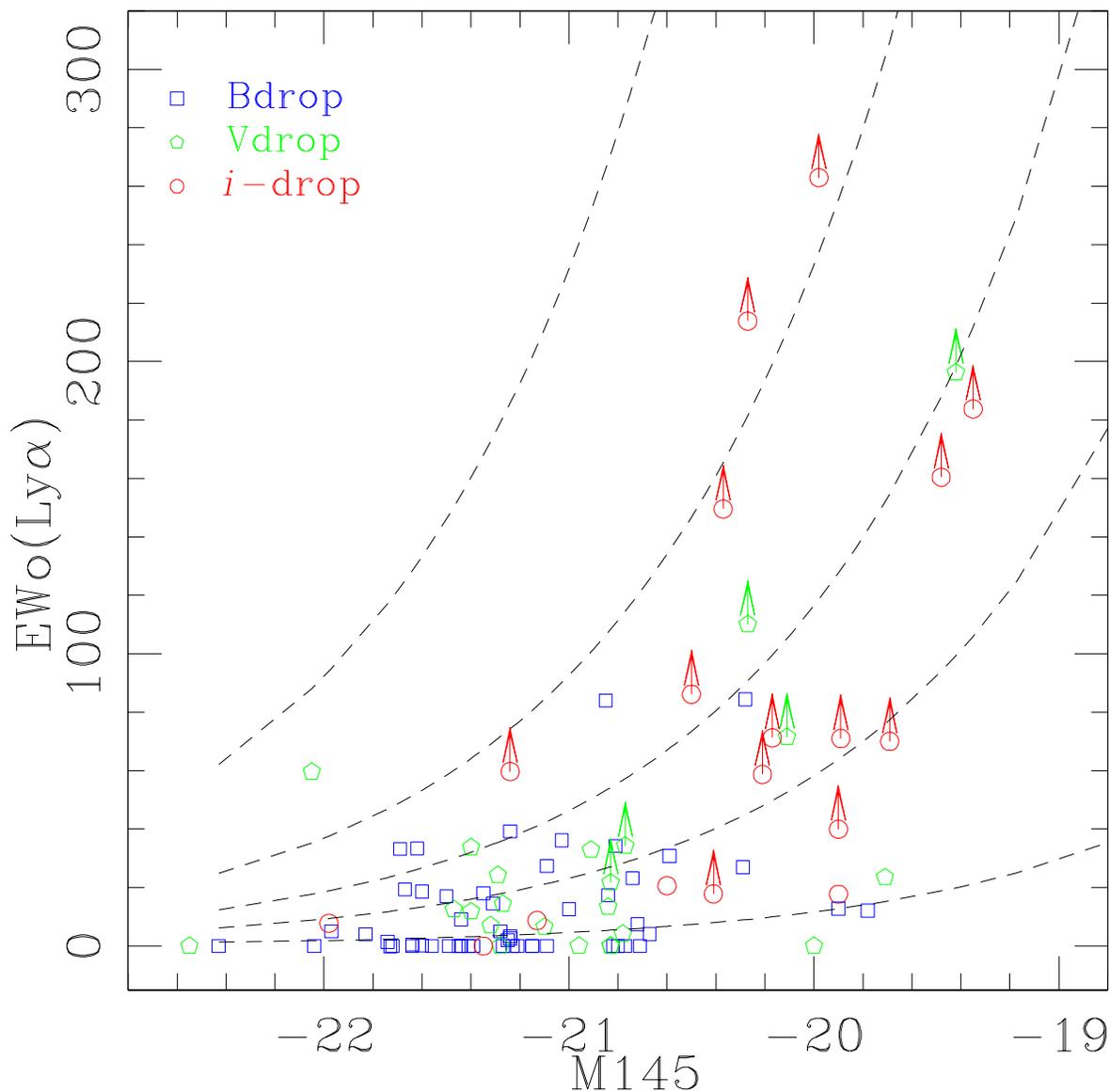} %EW_vs_M145.ps}
 \caption{Rest frame \Lya\ equivalent width as a function of the UV luminosity (M145, absolute magnitude at 1450\AA) for all the 
 galaxies at redshift beyond 3.1. Dashed curves represent the equivalent widths at fixed \Lya\ luminosity, from top
 to bottom 5, 2, 1, 0.5 and 0.1 $\times$ $10^{43}$ erg/s, respectively.  There is a clear trend for the \Lya\ 
 equivalent width to increase, on average and in its maximum value, for fainter objects. \label{EW_vs_M145}.}
\end{figure}

Figure~\ref{EW_vs_M145} shows the distribution of the rest--frame
equivalent widths versus the absolute magnitude calculated at
1450\AA~for all sources in the sample.  A cosmic time between 0.9
to 1.6 Gyr after the Big-Bang is covered (\wb, \wv\ and \wi\--band
dropouts are marked with different symbols).  At fainter luminosities
($M145 > -21$), the estimated equivalent widths span a wide range
of values, from a few Angstroms up to 300\AA . There is a natural
observational bias that the redshift of the faintest, high--redshift
galaxies can only be measured if they contain a strong, high
equivalent width \Lya\ line.  However, there is no such bias against
high equivalent width for {\it brighter} galaxies, but these are
not observed.  This absence of large equivalent widths of \Lya\
lines at bright luminosities has already been noted by several
groups studying samples of \Lya\ emitters (LAEs) and LBGs at redshift
between 3 and 6 \citep[e.g.,][]{shapley03, ando06, ando07, tapken07,
ver08}.

The equivalent width of the \Lya\ line (or the escape fraction of
the \Lya\ photons) is related to the velocity expansion $V_{\rm
exp}$ of the medium, the column density of the neutral gas $N_{\rm
HI}$, the dust extinction $E(B-V)$ and the geometry of the media
(clumpy or continuum geometry).  A possible scenario is that the
brighter galaxies are experiencing (or have already experienced) a
higher burst of star formation and supernovae explosions with an
associated production of dust.  Thus, the more luminous galaxies
would be dustier and more metal rich, have correspondingly more
efficient \Lya\ absorption, and thus exhibit lower observed \Lya\
equivalent widths.  Larger equivalent widths are expected for objects
dominated by younger ($\simlt 10-40$~Myr) stellar populations; lower
equivalent widths are expected in dusty and/or post-starburst
galaxies \citep[e.g.,][]{sch08}.  This hypothesis implies that
brighter LBGs would be dustier, more chemically enriched, and show
lower equivalent widths (\Lya).  One would expect that ultimately
the main underlying parameter governing the trends with UV magnitude
might be the galaxy mass.

Finally we note that fixing the redshift (i.e., the dropout flavor)
in Figure~\ref{EW_vs_M145}, the deficiency of strong lines at bright
UV magnitudes remains, though better statistics are clearly needed,
particularly at the faint end of the redshift distributions.

\begin{figure}
 \epsscale{1.}
 \plotone{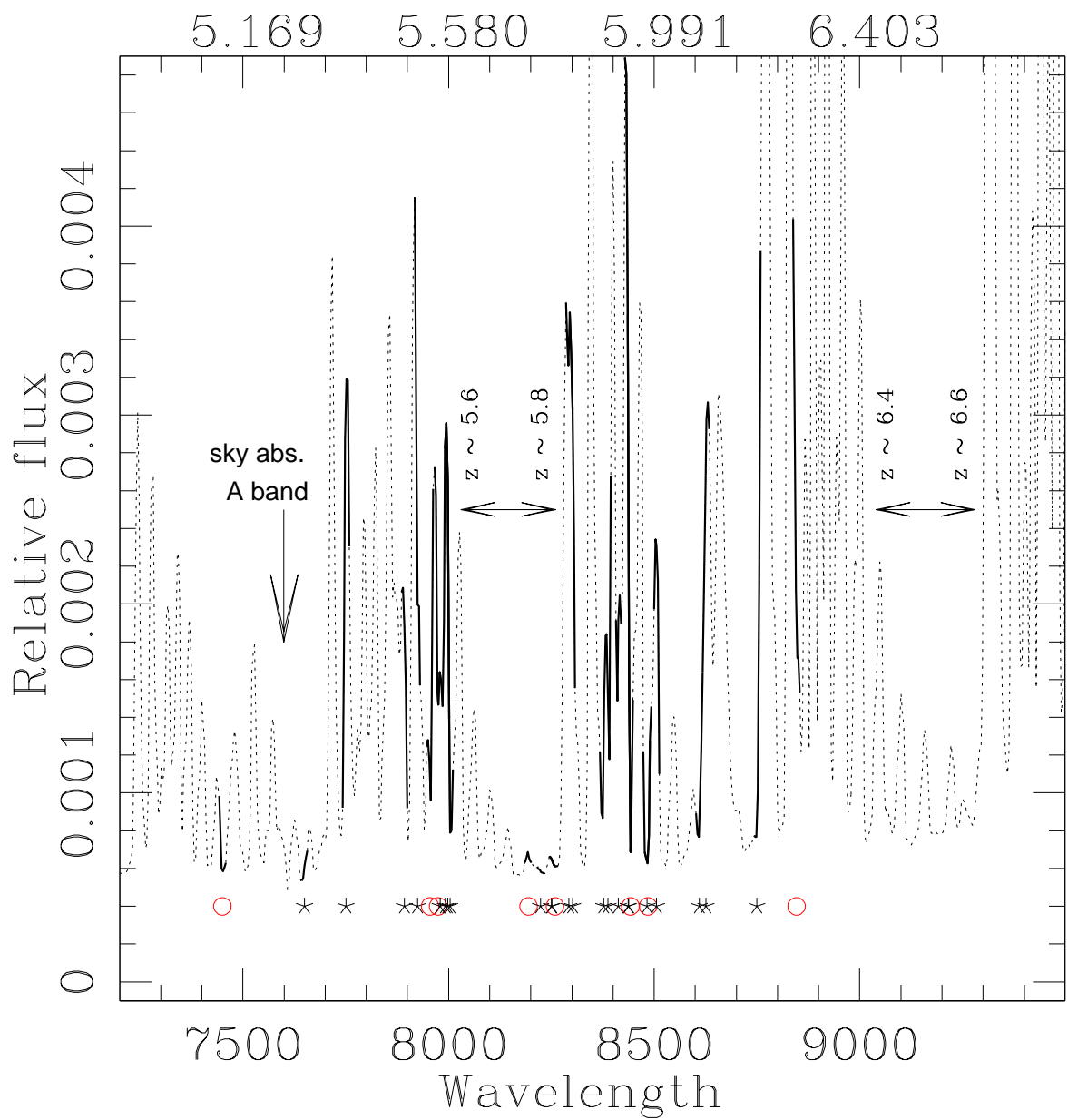} %LyA_and_SKY.ps}
 \caption{The positions of the \Lya\ lines (solid line) for galaxies at redshift beyond 5 are marked on the
 sky spectrum (dotted line). Stars and open circles denote the redshift of the \Lya\ position of
 galaxies with QF=A,B and QF=C, respectively.
 The lines have been detected sparsely in the forest of the sky emission.
 The sky free windows at redshift $\sim$ 5.7 and $\sim 6.5$ are also shown.  \label{LYA_SKY}}
\end{figure}

\begin{figure}
 \epsscale{1.}
 \plotone{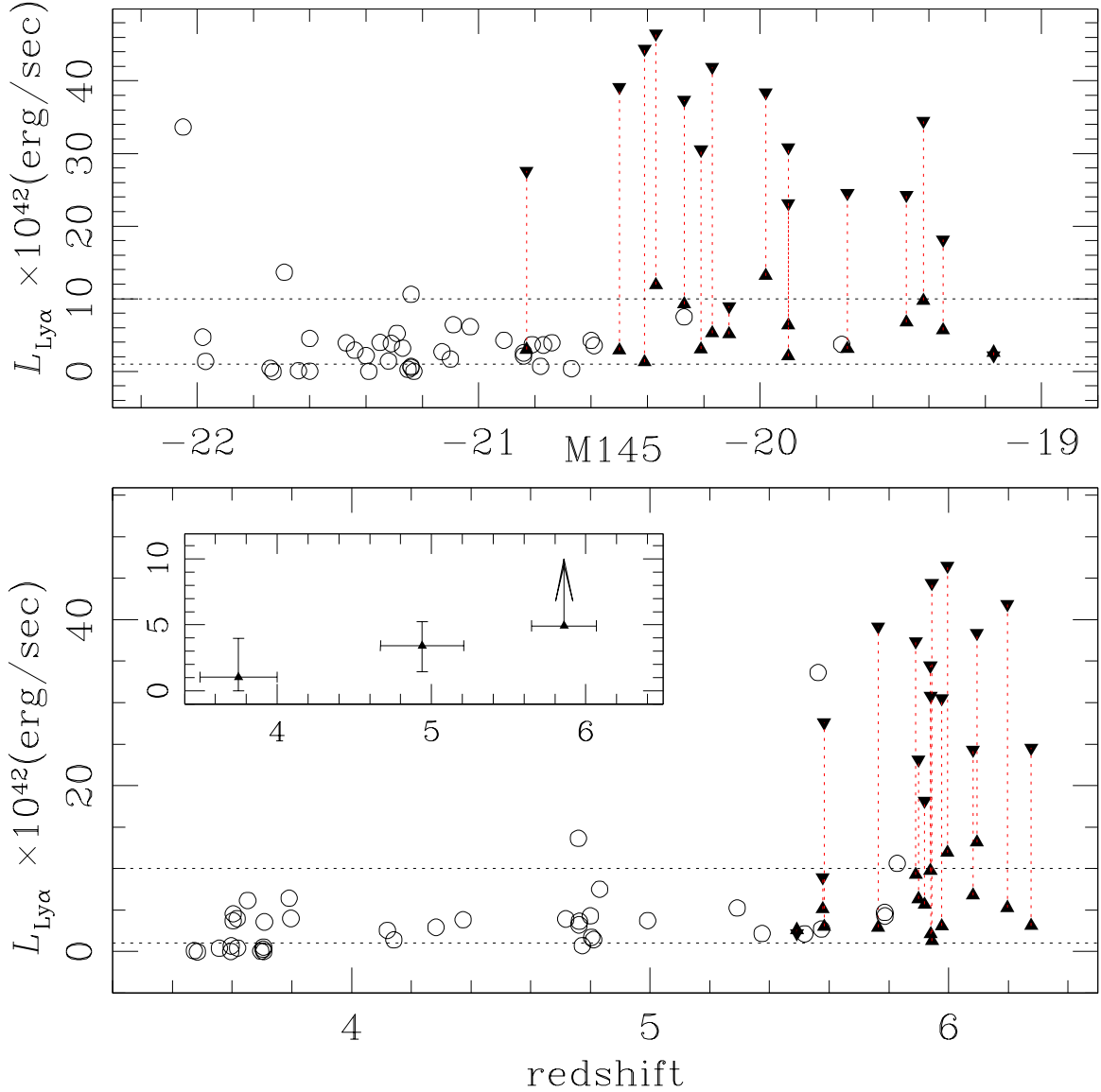} %LLyA_vs_M145_redshift.ps}
 \caption{{\bf TOP:} The \Lya\ luminosity as a function of the M145 magnitude. {\bf BOTTOM:} \Lya\ luminosity
as a function of the redshift. In the inner box, the same figure is shown but the medians have been
calculated at the average values of the three categories, \wb , \wv, \wi--band dropouts.
For both panels, the filled triangles connected by a dotted line represent the lower limit and upper
limit to the \Lya\ luminosity for galaxies without continuum detected in the spectra. 
The lower limit is simply the integral of the \Lya\ line, while the upper limit is
calculated assuming that the entire \wz\ flux is due to the line. 
Dotted horizontal lines mark the $10^{42}$ and $10^{43}$ erg/sec luminosity, respectively.  \label{LyA_lum}}
\end{figure}

\begin{figure}
  \epsscale{1.0}
  \plotone{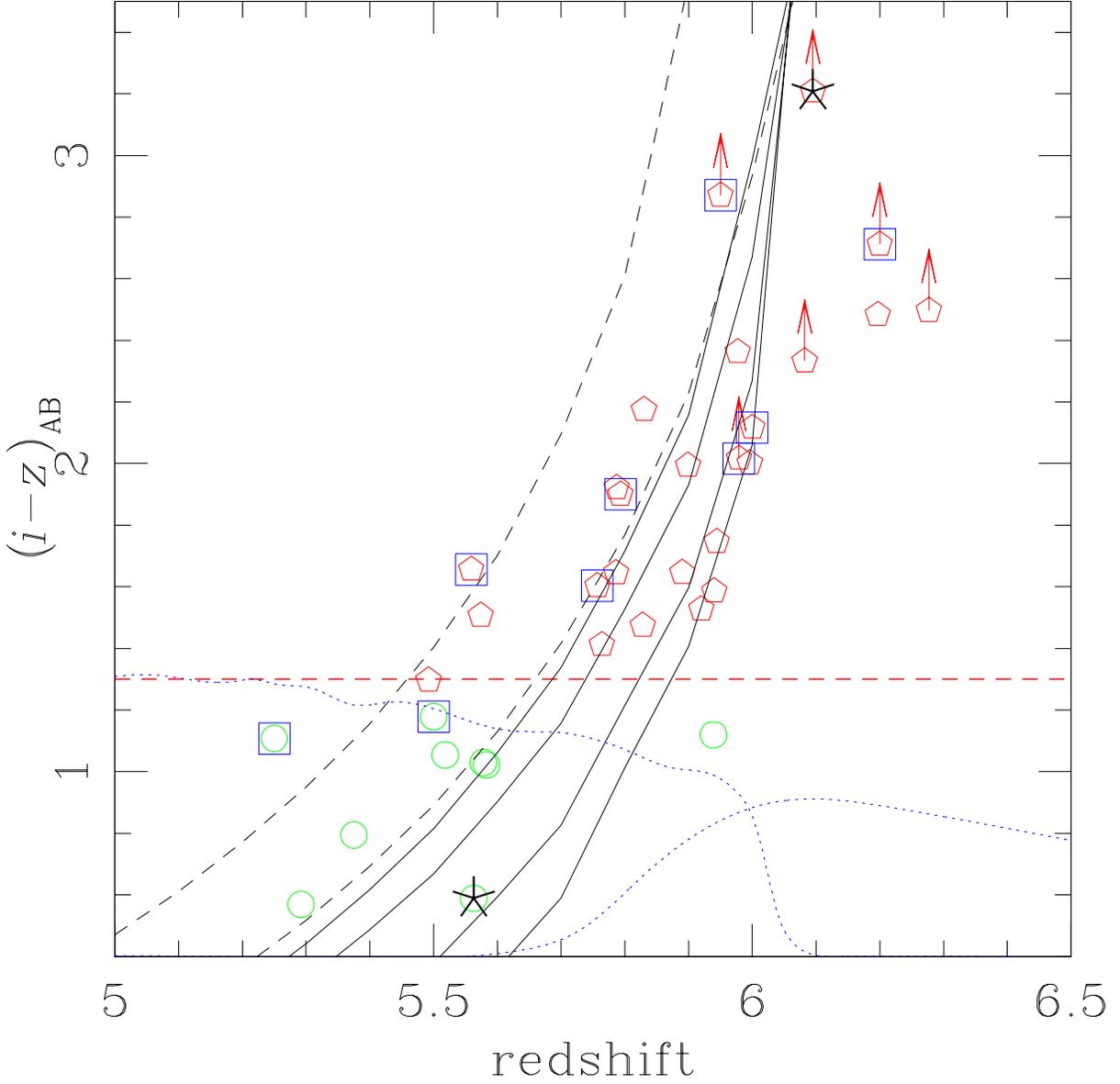} %izVSzspec.ps}
\caption{Color-redshift diagram of the spectroscopic sample
at redshift beyond 5. The six curves show the predicted
(\wi-\wz) color for different templates spectra. The templates
have been built combining synthetic spectra (drawn from SB99) with
different \Lya\ equivalent widths. The two dashed lines from left to right have
stellar populations of $10^{8}$ and $10^{7}$ years, respectively, and no
emission \Lya\ is present. The four solid lines from left to right
are color tracks assuming a fixed template of $10^{7}$ years (from SB99)
with the addition of \Lya\ emission line with rest frame equivalent widths of 
30, 50, 100, 150\AA, respectively. The attenuation of the intergalactic
medium has been implemented adopting the prescription of \cite{madau95}.
Open squares are sources with QF=C, pentagons and open circles are \wi\ and
\wv--band dropouts, respectively. The dotted curves are the shapes of the 
filters \wi\ and \wz , and show at which redshift the \Lya\ line enter 
and leave them. The two stars mark two peculiar 
galaxies (see text). \label{fig:izVSzspec}}
\end{figure}

\begin{figure}
 \epsscale{0.8}
 \plotone{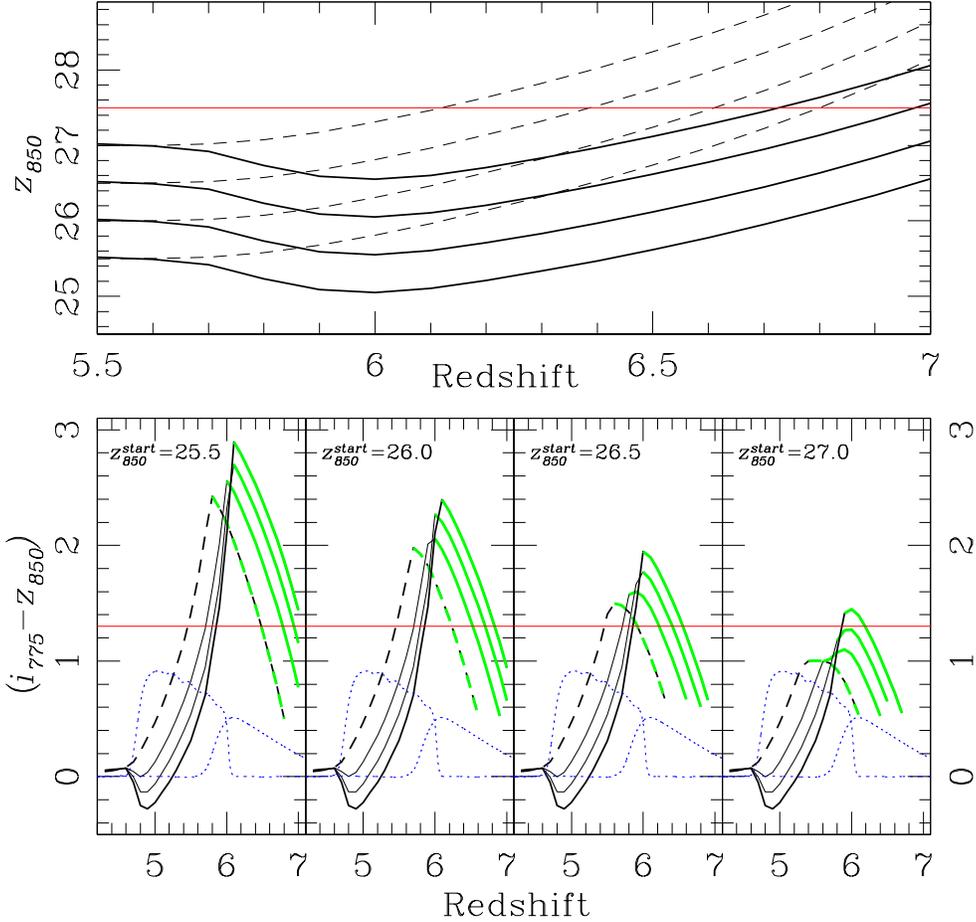} %SELTEST_IDROP_100Myr.ps} 
 \caption{{\bf Top:} Pure luminosity-redshift dimming of the \wz\ apparent
 magnitude calculated so that at redshift 5.6 (\Lya\ just blueward the \wz\ band) it is
 25.5, 26, 26.5, 27 from bottom to top curves. Solid line is the magnitude track
 of a SB99 template (with age of the stellar population of 100 Myr) with 
 \Lya\ emission inserted with rest frame equivalent width of 150\AA, the dashed lines without \Lya\ line
 inserted. {\bf Bottom:} The (\wi-\wz) color as a function of the
 redshift, the \wz\ magnitude and the \Lya\ rest frame equivalent width: dashed
 lines correspond to equivalent width of 0\AA, solid lines from left to right correspond 
 to 50, 100 and 150\AA, respectively.  The thick green lines represent the regions where the
 color becomes a lower limit (assuming the limit of the GOODS survey in the
 i775 band to be 28.0 at 2$\sigma$). All lines have been plotted with the
 condition \wz\ $<27.5$. The red horizontal line is the color cut adopted
 for the selection of z$\sim$6 galaxies. The shapes of the i775 and z850
 filters redshifted to the \Lya\ position are also reported, blue dotted
 lines (\Lya\ line enter the \wz\ band at redshift $\sim$ 5.6 and leaves the \wi\ band
 at redshift $\sim$ 5.9). It can be seen that fainter galaxies at $z>5.6$ tend to 
 be selected with strong \Lya\ line. \label{SELEZ_IDROP}}
\end{figure}

On the other side of the distribution, the presence of large \Lya\
equivalent widths for faint sources may be a combination of selection
effects and intrinsic properties of these galaxies:
\begin{enumerate}
\item{{\bf Observational bias:}
\begin{itemize} 
\item Spectroscopy.  Obviously, from the spectroscopic point of
view, faint galaxies (mainly \wi--band dropouts) are confirmed
thanks to the presence of a \Lya\ emission line that can be observed
also in the middle of the sky emission (see Figure~\ref{LYA_SKY}).
In the current spectroscopic sample, fainter galaxies tend to be
at higher redshifts.  Figure~\ref{LyA_lum} shows the behavior of
\Lya\ luminosity versus redshift. There is an indication that the
fraction of stronger lines increases with redshift.
\item Photometry.  Strong \Lya\ emission also affects photometric
color selection --- in particular, for \wi--band dropouts which
rely on a single color (i.e., \wi--\wz\ $>$ 1.3).  At $z > 5$, the
contribution of \Lya\ emission to the (\wi\--\wz) color range up
to $\approx$ 0.5 (0.8) magnitudes for \Lya\ rest--frame equivalent
widths of 100 \AA\, (150 \AA), consistent with the measurements in
our spectroscopic sample (see Figure~\ref{fig:izVSzspec}).  The
(\wi-\wz) color is increased or decreased depending on the strength
of the line and the redshift of the source.  Two clear examples
(marked with star symbols in Figure~\ref{fig:izVSzspec}) are:
\begin{enumerate}
\item{GDS~J033218.92-275302.7 ($z=5.563$) shows a \Lya\ emission
line with rest--frame equivalent width of $\sim 60$\AA\ falling
within the \wi\ band.  This has the effect of reducing the apparent
\wi--\wz\ color by 0.59 magnitudes.  This galaxy has been selected
as a \wv--band dropout and is also discussed as a candidate ``Balmer
Break galaxy'' based on its bright IRAC flux and apparent ``break''
in the $K-3.6\mu$m colors \citep{wik08}.}
\item{GDS~J033223.84-275511.6 ($z=6.095$) shows a \Lya\ emission
line with no continuum detected in 80~ks of spectroscopy.  The
rest--frame equivalent width is $\simgt 250$\AA, and the measured
(\wi-\wz) color is a lower limit ((\wi-\wz ) $> 3.2$).  In this
case the \wz\ apparent magnitude (and the (\wi-\wz) color) is
increased by the line.}
\end{enumerate}
In order to explore such effects as a function of redshift, \Lya\
equivalent width and \wz\ magnitude, we have calculated various
color tracks as shown in Figure~\ref{SELEZ_IDROP}.  We find that,
when the \Lya\ line enters the \wz\ band ($z > 5.6$) and leaves the
\wi\ band ($z > 5.9$), depending on the equivalent width, it favors
the \wi\--band dropout selection criteria.  For fainter sources
(\wz$>$26.5), only the emitters tend to survive.
\end{itemize}
}
\item{{\bf Intrinsic effects:} the large spread in \Lya\ equivalent
widths at faint magnitudes ($M145 \simlt -21$) observed by numerous
authors may also be due to a relatively small amount of dust, which
would not filter out the stronger \Lya\ lines, and to a larger
variety of star formation histories and timescales --- i.e., an
enhanced role of ``stochastic star formation events.'' Such a
scenario is most likely to have a strong effect for galaxies of
smaller absolute scale \citep[either mass or total star formation
rate,][]{ver08}.}
\end{enumerate}

We further note that at $z \simgt 6$, the age of the Universe is
of the order of the duration of the LBG phase 
\citep[$\sim 0.5-1$ Gyr ---][]{shapley01, papovich01, lee08}.

Assuming an initial interval of time ($\Delta t_{\rm Ly\alpha}$)
in which the LBG is active as a LAE (i.e., shows conspicuous \Lya\
emission, with rest--frame equivalent width greater than 100\AA),
whose duration should be of the order of $100-300$ Myr
\citep[e.g.,][]{mori06, ver08}, the probability to observe a LBG
in the LAE phase should increase with redshift when observing
galaxies in a Universe younger than $\sim 0.5-1$ Gyr (roughly, the
fraction of emitters versus non-emitters is proportional to $\Delta
t_{\rm Ly\alpha} / \tau(z)$).  Future surveys of LBGs at redshift
beyond seven should show this trend even more clearly (albeit subject
to the observational selection effects discussed above).

\section{Correlation with Morphological Properties}

We have derived basic morphological parameters for the galaxies in
our spectroscopic samples from the ACS \wz--band image. With an
effective wavelength $\lambda_{\rm eff}\approx 9100$ \AA\ (for a
typical LBG UV spectrum), the \wz\ filter probes the rest--frame
far--UV emission of \wb--dropout galaxies at $\lambda_0 \approx
2000$ \AA. In general, it is difficult to interpret the results of
analyses of the UV morphologies of high--redshift galaxies in terms
of the evolution of traditional Hubble types, in part because these
are mostly known at optical rest--frame wavelengths (c.f., Giavalisco
et al.\ 1996a,b), and also because it is not obvious what is the
typical morphology of the present--day spectral types that are most
similar to the $z \sim 4$ LBGs.

We have measured parametric and non--parametric morphological
indicators separately for the two sub--samples of ``emitters'' and
``absorbers'' in the $z\sim 4$ primary sample.  The basic morphological
parameters have been drawn from the v2.0 ACS catalogs, direct outputs
of the SExtractor algorithm during the segmentation process in the
\wz\ band, and are summarized in Table~\ref{morph} with their average
values and 1-$\sigma$ standard deviations.  Tabulated quantities
are the major semi-axis ($a$), half--light radius ($h.l.r.$),
isophotal $area$ (AREAF) and FWHM.  As shown in Table~\ref{morph},
LBGs with \Lya\ in emission have more compact morphologies relative
to those with rest--frame UV features observed in absorption.  In
detail, the physical sizes at the half light radius for emitters
and absorbers are on average 1.1 and 1.6 kpc, respectively.

To further investigate the correlation between morphology and \Lya\
properties, we have computed the Gini coefficient for our sample.
We have utilized the formulation described in \citet[][, their
Eq.~3]{abra03}.  The Gini coefficient provides a measure of the
degree of central concentration of the source. Values range between
0 (uniform surface brightness) and 1 (highly nucleated).  \cite{lotz06},
\cite{ravi06} and, more recently, \cite{lisk08}, have analyzed the
stability of the Gini coefficient, based on a comparison of {\it
HST}/ACS imaging data from the GOODS and UDF surveys.  They find
the Gini coefficient depends strongly on the S/N ratio and at all
S/N levels, the Gini coefficient shows a strong dependence on the
choice of aperture within which it is measured.  This complicates
comparisons of Gini parameters derived in different studies.  However,
relative values from measurements done the same way within a given
data set should be meaningful.  In the present case we restrict the
analysis for the brighter \wb--dropout sample and assume that
systematics are similar for both emitter-- and absorber--class LBGs.
The pixels of each source used in the calculation are those with
flux above F $\times$ 1$\sigma$ percentile of the median background.
Adopting F=2 and the \wz\ band (F=2, \wz\ band), we find that the
``em.'' and ``abs.'' classes have $G_{\rm em}$=$0.41^{+0.11}_{-0.06}$
and $G_{\rm abs}$=$0.26^{+0.18}_{-0.10}$, respectively. With (F=3,
\wz\ band) the values are $G_{\rm em}$=$0.31^{+0.09}_{-0.09}$ and
$G_{\rm abs}$=$0.18^{+0.11}_{-0.08}$.  The same calculation performed
in the \wi\ band, produces the following median values: $G_{\rm
em}$=$0.49^{+0.10}_{-0.16}$, $G_{\rm abs}$=$0.26^{+0.14}_{-0.09}$
(F=3, \wi\ band) and $G_{\rm em}$=$0.61^{+0.11}_{-0.15}$,  $G_{\rm
abs}$=$0.35^{+0.19}_{-0.14}$ (F=2, \wi\ band).

\begin{figure}
 \epsscale{0.8}
 \plotone{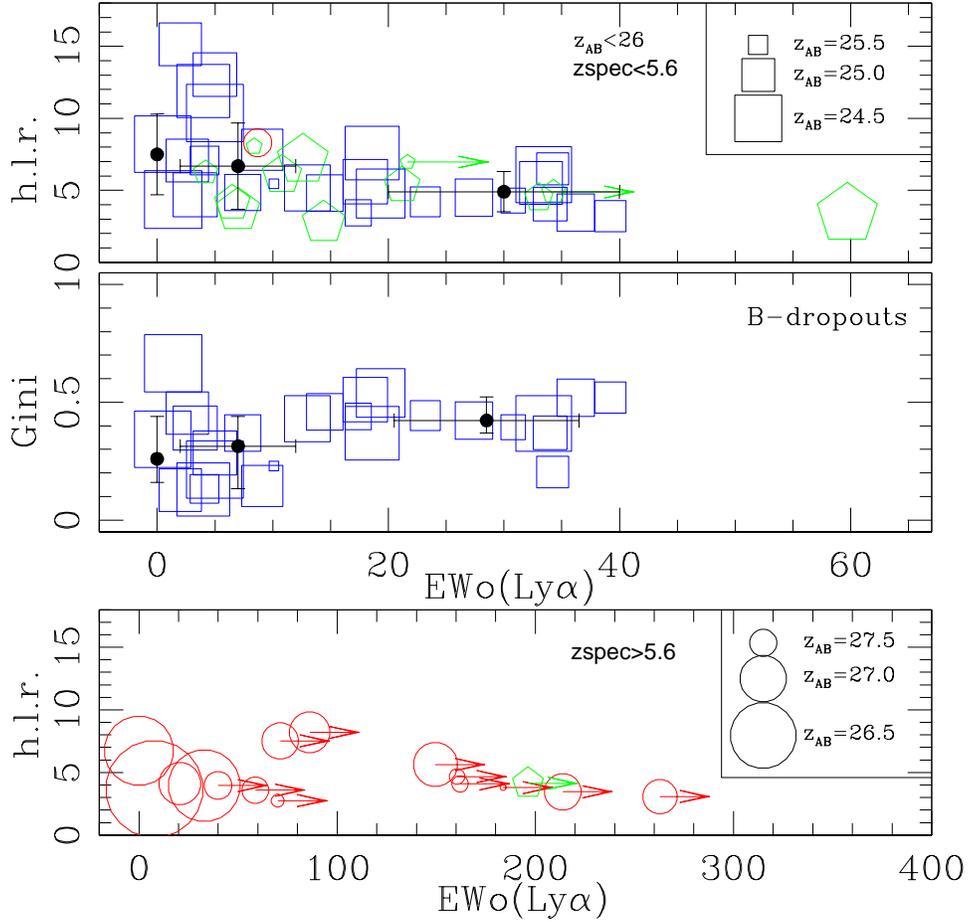} %MORPH.ps} 
 \caption{{\bf Top:} Half light radius versus \Lya\ equivalent width. Squares, pentagons and
open circles mark \wb\ , \wv\ and \wi--dropouts, respectively. The size of the symbols scale
with apparent \wz\ magnitude. Only sources with \wz\ below 26 and redshift below 5.6 
($\sim$ 0.8 Gyrs of the cosmic time is probed in the redshift range 3.5-5.6) are
plotted (see text for details). For comparison, the filled circle at 0\AA~is the average 
half light radius for galaxies without \Lya\ emission line. The other two filled circles
are the averages for sources in the bin 0-20\AA~and beyond 20\AA. 
There is an apparent behavior such that larger \Lya\
equivalent widths corresponds to smaller galaxies. {\bf Middle:} Gini coefficient versus 
\Lya\ equivalent width only for the \wb--dropout sample. The filled circle at 0\AA~ is
the average of the absorbers, the other two filled circles are the averages of the Gini 
parameter in the bin 0-20\AA~and beyond 20\AA. Sources with
larger \Lya\ equivalent widths seems to be more nucleated. {\bf Bottom:} Same of top panel, but for
galaxies with redshift beyond 5.6 (mainly \wi--dropouts). In all panels
no QF=C have been considered.   \label{MORPH}}
\end{figure}

These calculations show that the two LBG spectroscopic classes have
different average morphologies, with emitters intrinsically more
nucleated than the absorbers. This distinction seems to increase
with greater \Lya\ equivalent width.  The behavior is shown in
Figure~\ref{MORPH} (middle panel, F=2, \wz\ band), where the Gini
coefficient is plotted versus the \Lya\ equivalent width.  Though
this result is, on average, in qualitative agreement with the
observations at $z\sim$ 2 and 3 by \citet{law07}, we note that cases
of nucleated absorbers and ``fuzzier'' emitters are also present.
Larger galaxy samples at these redshifts are needed in order to put
this result on a firmer statistical footing.

As shown in Figure~\ref{MORPH} (top panel), there also seems to be
an inverse correlation between \Lya\ emission equivalent width and
galaxy size, namely galaxies with larger equivalent widths are
smaller.  To some extent, this correlation can be explained by the
fact that galaxies with larger equivalent widths are more likely
to be fainter; this is the case for the \wi--dropout sample
(Figure~\ref{MORPH}, bottom panel).  The correlation, however, seems
to persist even when subsamples cut by absolute luminosities are
considered, as illustrated in the top panel of Figure~\ref{MORPH}
where the size of the symbols scale with apparent \wz\ magnitude.
In this latter case, only sources with spectroscopically-detected
detected and $z < 5.6$ have been considered (at $z> 5.6$, \Lya\
enters the \wz\ band). One potential physical explanation of this
size behavior could lie in the masses of the objects. \Lya--emitting
LBGs at $z \sim 3-4$ are found to have smaller stellar masses than
objects lacking this emission \citep[e.g.][]{gawiser06}.  Pentericci
et al.\ (2007) also found for the \wb--band dropout sample in the
present work an average stellar mass of $5 \pm 1 \times  10^9$
M$_{\odot}$ and $2.3 \pm 0.8 \times 10^{10}$ M$_{\odot}$ for the
emitters and absorbers, respectively. This further suggests that
emitters may be associated with less massive dark matter halos and
hence have experienced a different star formation history compared
to the absorption line galaxies.  At higher redshift, this analysis
is more critical because high quality near--infrared images (e.g.,
NICMOS) and deep spectroscopy are needed to identify the absorbers.
\citet{hyge07} found that \Lya--emitting \wi--band dropouts seem
to be morphologically distinct from the general \wi--band dropout
LBGs.  We only report here that, on average, the $h.l.r.$ for our
sample of \wi--dropouts emitters is consistent with other observations
\citep[e.g.][]{stan04a, stan04b, hyge07}, with $h.l.r. \sim 0\farcs13$
(Figure~\ref{MORPH}, lower panel).  The only \wi--band dropout with
QF=B without \Lya\ in emission (GDS~J033233.19-273949.1) has $h.l.r.
= 0\farcs.20$.  High quality and deeper near-IR images and spectroscopy
are necessary to investigate this issue.

\section{Conclusions} 

In the present work, we have addressed the spectroscopic properties
of LBGs at high redshift, selected from the GOODS survey.  We have
discussed the efficiency of the photometric selection criteria
adopted.  We have extracted preliminary information from the spectral
features and UV luminosity and compared it with analogous studies
at lower redshift. Summarizing,
\begin{enumerate}
\item{109 out of 202 targeted LBGs have been spectroscopically
confirmed in the redshift range $3.1 < z < 6.6$, according to
\wb--,\wv-- and \wi--band dropout selections.  This relatively low
confirmation rate is largely due to the following two reasons:  i)
the target list includes a relatively large fraction of faint
sources, with 65 out of 202 or 32.2\% of the sample having \wz$>26$;
and ii) the difficulty in determining redshifts for galaxies at $z
< 3.6$ given our instrumental set-up.  Considering sources with
determined redshifts, 96$\%$, 89$\%$ and 82$\%$ of the observed
\wb--,\wv-- and \wi--band dropout samples have been confirmed in
the expected redshift range, respectively.  Twelve low--redshift
interlopers have also been confirmed, 10 stars and two galaxies at
$z < 2$.  Five high--redshift galaxies have been serendipitously
discovered, yielding a total of 114 redshifts measured beyond
redshift 3.1 (38 of these with QF=C).}

\item{From the composite spectra of the three flavors of dropout
(\wb, \wv\ and \wi--band dropouts), we detect the typical spectral
features of star--forming galaxies, namely a flat spectrum redward
of \Lya, IGM attenuation and the Lyman limit blueward of \Lya, UV
absorption lines (both high and low ionization), and \Lya\ seen in
both emission or absorption.  In particular, at $z \sim 4$, a
comparison between the composite spectra of emitters and absorbers
shows steeper spectral slopes and weaker UV absorption features for
the emitters.}

\item{Galactic outflows have been identified at $z \sim 4$ by
measuring the velocity offset between interstellar, \Lya\ and nebular
lines.  The measured $\langle V_{\rm Ly\alpha} - V_{\rm ISL} \rangle
= 370_{-116}^{+270}$ \kms\ is consistent with results at $z \sim
3$ by \citet{shapley03}, considering the portion of their sample
with similar \Lya\ equivalent widths to our sample.  We derive
$\langle V_{\rm ISL} - V_{\rm nebular} \rangle$ of $-165$ \kms,
similar to the $-150$ \kms\ derived by \citet{adel03} at lower ($z
\sim 3$) redshift.  A similar offset (but less accurate because it
is derived from the composite spectrum) has been detected in the
\wv--dropout sample (redshift $\sim$5), i.e., $\langle V_{\rm
Ly\alpha} - V_{\rm ISL} \rangle \sim 500$ \kms.  This supports the
interpretation that outflows similar to those taking place at $z
\sim$ 2 and 3 are also observed in our samples of LBGs at $z\sim$
4 and 5.}

\item{The presence of a weaker \Lya\ equivalent widths for dropouts
with brighter UV luminosities ($M_{145} < -21$) is clear in the
current spectroscopic sample (considering all categories).  This
trend has been recently noted by several authors, and may be naturally
explained by a different evolution of bright UV LBGs with respect
to the fainter ones. The brighter galaxies should be dustier and
more evolved (and probably more massive) than the fainter ones,
which show a larger spread of \Lya\ equivalent widths possibly due
to assorted SF histories.}

\item{The sample at $z \sim 4$ exhibits correlations between certain
basic UV rest--frame morphological properties and spectroscopic
properties such as the presence and strength of \Lya\ emission.  In
particular, emitters appear more compact and nucleated than absorbers.
\citet{law07} find a similar ``nucleation effect'' at $z \sim$ 2
and 3 in their BM/BX and LBG samples, and interpret this as a
consequence of more dust in the absorbers leading to redder colors
and more diffuse morphologies.  Pentericci et al.\ (2008) analyze
the photometric properties of the same sample as discussed here,
and find that emitters are less massive and less dusty than absorbers.
Focusing on the emitters, increasing \Lya\ equivalent widths
correspond to decreasing stellar masses and extinction.  The emitters,
especially those with a large \Lya\ equivalent widths, could be
systems forming a relatively large fraction of their stellar mass
during an intense burst of star formation.  These putative
proto--spheroids observed at $z \sim 4$ could include in significant
numbers the progenitors of the compact massive early--type galaxies
identified at $z \sim 2$ (e.g., Cimatti et al.\ 2008; van Dokkum
et al. 2008; Buitrago et al. 2008).  Such an evolutionary link is
generally consistent with the observed spatial clustering properties
and the stellar populations of LBGs at $z \sim 3$ and $\sim 4$
\citep{giava98, giava01, lee06, lee08, ouchi05} and those of the
BzK and DRG galaxies at $z \sim 2 -2.5$ \citep{kong06, quadri08}.
The strength of spatial clustering increases with the mass of the
galaxies and with redshift, as a consequence of gravitational
evolution of structure. The observed larger spatial correlation
length and larger stellar mass of the UV/optical--selected galaxies
at $z\sim 2$ \citep{daddi07, vandokkum06} are in overall quantitative
agreement with the expected dependence of clustering with both mass
and time, when compared to the less strongly clustered and less
massive (in stellar content) UV--only selected galaxies at $z\sim
3$ and $\sim 4$ \citep{adel05}. This suggests that the same populations
of dark matter halos is being observed at different evolutionary
stages of the growth of their galaxy hosts and spatial clustering.}

\end{enumerate}

\acknowledgments
We are grateful to the ESO staff in Paranal and Garching who greatly
helped in the development of this programme. We acknowledge financial
contribution from contract ASI/COFIN I/016/07/0 and PRIN INAF 2007
``A Deep VLT and LBT view of the Early Universe''. EV
thanks STScI and NOAO for hospitality during a visit in which this paper 
was conceived and partially written. EV thanks F. Calura for useful 
discussions about the dust properties of high redshift galaxies. 
The work of DS was carried out at Jet Propulsion Laboratory, California 
Institute of Technology, under a contract with NASA.

%%%%%%%%%%%%%%%%%%%%%%%%% 
\clearpage
\begin{deluxetable}{lccccccccc}
\tabletypesize{\scriptsize}
\tablecaption{The spectroscopic sample of the \wb--band dropouts. 
The redshift reported is the result of the cross-correlation between the
spectrum and the reference template. In the first four columns the GDS name, the
redshift value, QF and class, are listed, respectively. Columns $\#$5 to $\#$10 are the
\wz\ AB magnitude (MAGAUTO), the half light radius, the galaxy-star classifier 
(0=galaxy, 1=point-like source), the (\wb--\wv) and the (\wv--\wz) colors, and
the isophotal S/N ratio in the \wb\ band (if $<$1, the (\wb--\wv) color is
a lower limit).  \label{tab:Bdrops}}
\tablewidth{0pt}
\tablehead{
\colhead{GOODS ID} & \colhead{$z$} & \colhead{QF} & 
\colhead{class} & \colhead{\wz\ } & \colhead{h.l.r.} & 
\colhead{S/G} & \colhead{(\wb\--\wv\ )} & \colhead{(\wv\--\wz\ )} 
& \colhead{$(S/N)_{B}$}}
\startdata
J033200.31-274250.7 & 0.000 &B &star & 23.43 & 2.65 & 0.99 & 2.27 & 0.72 & 15.72\\
J033239.12-274751.6 & 1.541 &B &em. & 25.49 & 4.14 & 0.16 & 2.59 & 1.42 & 1.02\\
J033242.84-274702.5~\tablenotemark{a} & 3.193 &B &em. & 24.92 & 3.07 & 0.96 & 1.37 &-0.01 & 17.00\\
J033234.83-275325.2 & 3.369 &B &abs. & 24.24 & 9.65 & 0.03 & 1.82 & 0.62 & 7.19\\
J033220.85-275038.9 & 3.450 &B &abs. & 24.56 & 14.09 & 0.00 & 2.02 & 0.58 & 3.73\\
J033223.34-275156.9 & 3.470 &A &abs. & 23.35 & 5.52 & 0.03 & 2.00 & 0.75 & 12.30\\
J033223.22-275157.9 & 3.470 &A &abs. & 25.07 & 6.75 & 0.03 & 1.50 & 0.33 & 7.31\\
J033214.82-275204.6 & 3.473 &A &comp. & 24.14 & 8.19 & 0.03 & 1.53 & 0.40 & 16.71\\
J033235.06-275234.6 & 3.477 &C &comp. & 25.12 & 8.28 & 0.03 & 1.61 & 0.43 & 6.15\\
J033220.97-275022.3 & 3.478 &A &abs. & 24.70 & 8.86 & 0.03 & 1.82 & 0.44 & 5.84\\
J033225.16-274852.6 & 3.484 &A &comp. & 24.05 & 6.73 & 0.03 & 1.58 & 0.45 & 14.76\\
J033223.99-275216.1 & 3.557 &B &comp. & 25.15 & 7.06 & 0.03 & 1.59 & 0.40 & 6.81\\
J033226.76-275225.9 & 3.562 &A &abs. & 24.10 & 6.69 & 0.03 & 1.69 & 0.49 & 12.64\\
J033229.02-274234.0 & 3.585 &B &abs. & 25.01 & 5.66 & 0.03 & 1.66 & 0.21 & 10.59\\
J033220.94-274346.3~\tablenotemark{\star} & 3.596 &A &em. & 24.61 & 6.37 & 0.45 & 1.75 & 0.34 & 9.53\\
J033229.14-274852.6 & 3.597 &A &em. & 24.60 & 4.69 & 0.03 & 1.70 & 0.34 & 11.64\\
J033201.84-274206.6 & 3.603 &A &em. & 25.04 & 6.50 & 0.01 & 1.71 & 0.07 & 9.50\\
J033242.50-274551.7 & 3.604 &A &em. & 24.24 & 7.59 & 0.04 & 1.75 & 0.27 & 9.35\\
J033217.13-274217.8 & 3.617 &A &em. & 25.11 & 4.20 & 0.36 & 1.70 & 0.19 & 9.11\\
J033235.96-274150.0 & 3.618 &A &comp. & 24.11 & 4.36 & 0.03 & 1.65 & 0.39 & 16.16\\
J033215.78-274145.6 & 3.646 &C &abs. & 24.79 & 5.69 & 0.48 & 1.92 & 0.46 & 6.47\\
J033217.22-274754.4~\tablenotemark{\star} & 3.652 &A &em. & 24.84 & 3.45 & 0.33 & 1.77 & 0.26 & 11.10\\
J033222.59-275118.0 & 3.660 &A &abs. & 25.10 & 6.49 & 0.03 & 2.20 & 0.47 & 4.25\\
J033245.57-275333.3 & 3.685 &A &abs. & 24.61 & 4.90 & 0.03 & 1.83 & 0.64 & 7.02\\
J033217.66-275332.0~\tablenotemark{\star} & 3.696 &B &em. & 24.29 & 5.43 & 0.02 & 2.29 & 0.54 & 7.27\\
J033232.08-274136.4 & 3.697 &B &abs. & 24.74 & 7.30 & 0.03 & 2.49 & 0.56 & 3.68\\
J033230.10-275057.7 & 3.704 &A &comp. & 24.64 & 15.14 & 0.02 & 1.95 & 0.43 & 3.78\\
J033226.28-275245.7 & 3.705 &B &comp. & 24.65 & 7.07 & 0.32 & 2.01 & 0.30 & 6.93\\
J033218.05-274519.0 & 3.706 &A &abs. & 24.61 & 13.03 & 0.02 & 3.79 & 0.68 & 0.96\\
J033219.81-275300.9 & 3.706 &A &comp. & 24.50 & 5.82 & 0.03 & 1.95 & 0.54 & 8.10\\
J033219.60-274840.0 & 3.708 &A &em. & 25.30 & 4.28 & 0.40 & 1.93 & 0.36 & 5.67\\
J033225.82-274250.3 & 3.770 &C &abs. & 25.00 & 7.43 & 0.03 & 3.48 & 0.70 & 1.01\\
J033233.33-275007.4 & 3.791 &A &em. & 24.84 & 4.50 & 0.05 & 2.20 & 0.29 & 5.50\\
J033234.65-274115.4 & 3.794 &C &abs. & 24.62 & 7.66 & 0.02 & 2.43 & 0.70 & 2.54\\
J033236.83-274558.0 & 3.797 &A &comp. & 24.58 & 5.59 & 0.29 & 2.10 & 0.61 & 5.54\\
J033239.67-274850.6~\tablenotemark{\star} & 3.887 &B &abs. & 24.56 & 5.15 & 0.03 & 3.05 & 1.10 & 1.57\\
J033238.73-274413.3 & 4.000 &C &abs. & 24.81 & 11.18 & 0.00 & 2.95 & 0.96 &-0.21\\
J033227.94-274618.6 & 4.000 &C &abs. & 25.23 & 4.23 & 0.03 & 3.21 & 1.10 &-1.38\\
J033246.25-274847.0 & 4.020 &A &abs. & 24.88 & 5.09 & 0.03 & 3.71 & 0.77 & 0.82\\
J033241.16-275101.5 & 4.058 &B &abs. & 25.25 & 7.20 & 0.03 & 3.21 & 0.83 & 0.16\\
J033240.38-274431.0 & 4.120 &A &em. & 25.24 & 3.45 & 0.64 & 3.07 & 0.48 & 2.20\\
J033234.35-274855.8 & 4.142 &A &comp. & 24.11 & 10.37 & 0.03 & 3.11 & 1.02 & 1.66\\
J033218.26-274802.5 & 4.280 &A &abs. & 24.65 & 4.35 & 0.05 & 3.62 & 1.08 &-0.75\\
J033212.98-274841.1~\tablenotemark{\star} & 4.283 &B &em. & 24.70 & 7.81 & 0.03 & 3.63 & 0.69 &-0.29\\
J033248.24-275136.9 & 4.374 &A &em. & 24.87 & 4.81 & 0.09 & 3.37 & 1.11 &-0.54\\
J033214.50-274932.7 & 4.738 &C &em. & 25.40 & 5.30 & 0.30 & 3.00 & 0.96 &-0.08\\
J033257.17-275145.0 & 4.760 &A &em. & 24.64 & 5.52 & 0.02 & 2.71 & 1.53 & 1.47\\
\tableline
\enddata
\tablenotetext{a}{It has been identify with broad MgII in emission, QSO.}
\tablenotetext{\star}{For these sources, the QF has been improved (i.e. C to B or B to A) to respect 
the online release of \cite{vanz08}, after re-analyzing the whole LBG sample.}
\end{deluxetable}

%%%%%%%%%%%%%%%%%%%%%%%%%
\clearpage
\begin{deluxetable}{lccccccccc}
\tabletypesize{\scriptsize}
\tablecaption{The spectroscopic sample of the \wv--band dropouts. Cloumns as described in
Table~\ref{tab:Bdrops}.\label{tab:Vdrops}}
\tablewidth{0pt}
\tablehead{
\colhead{GOODS ID} & \colhead{$z$} & \colhead{QF} & 
\colhead{class} & \colhead{\wz\ } & \colhead{h.l.r.} & 
\colhead{S/G} & \colhead{(\wv\--\wi\ )} & \colhead{(\wi\--\wz\ )} 
& \colhead{$(S/N)_{V}$}
}
\startdata
J033242.08-274911.6 & 0.000 &B &star & 23.43 & 2.88 & 0.98 & 2.63 & 1.25 & 6.34\\
J033224.11-274102.1 & 0.000 &A &star & 23.39 & 2.61 & 0.99 & 2.45 & 0.87 & 12.10\\
J033237.69-275446.4 & 0.000 &A &abs. & 23.60 & 2.68 & 0.99 & 2.40 & 1.04 & 9.42\\
J033220.31-274043.4 & 1.324 &B &abs. & 24.09 & 10.18 & 0.00 & 2.10 & 1.16 & 3.40\\
J033242.62-275429.0 & 4.400 &C &abs. & 25.61 & 6.60 & 0.02 & 2.03 & 0.31 & 4.11\\
J033222.88-274727.6 & 4.440 &B &abs. & 24.92 & 4.05 & 0.03 & 1.63 & 0.09 & 14.73\\
J033222.97-274629.1 & 4.500 &C &abs. & 25.34 & 6.91 & 0.02 & 1.72 & 0.19 & 6.63\\
J033228.56-274055.7 & 4.597 &B &abs. & 25.44 & 7.53 & 0.00 & 1.58 & 0.02 & 9.29\\
J033216.98-275123.2 & 4.600 &B &abs. & 25.30 & 6.56 & 0.03 & 1.68 &-0.08 & 9.02\\
J033255.08-275414.5 & 4.718 &A &em. & 24.83 & 7.11 & 0.01 & 2.40 & 0.15 & 5.77\\
J033247.58-275228.2 & 4.758 &C &em. & 25.73 & 8.33 & 0.01 & 1.87 & 0.14 & 4.28\\
J033229.29-275619.5~\tablenotemark{a} & 4.762 &A &em. & 25.05 & 2.76 & 0.99 & 1.65 & 0.15 & 12.95\\
J033243.53-274919.2 & 4.763 &A &em. & 25.56 & 4.89 & 0.02 & 2.13 & 0.12 & 5.13\\
J033240.12-274535.5 & 4.773 &B &em. & 25.55 & 6.23 & 0.02 & 1.62 & 0.06 & 7.24\\
J033221.93-274533.1 & 4.788 &C &abs. & 25.82 & 4.86 & 0.04 & 2.17 & 0.23 & 3.56\\
J033228.85-274132.7 & 4.800 &B &em. & 25.43 & 4.50 & 0.03 & 1.66 &-0.02 & 9.84\\
J033205.26-274300.4 & 4.804 &A &em. & 25.24 & 4.11 & 0.03 & 1.85 &-0.04 & 11.39\\
J033210.03-274132.7 & 4.811 &A &em. & 25.03 & 3.63 & 0.31 & 1.77 & 0.12 & 12.63\\
J033242.66-274939.0 & 4.831 &B &em. & 26.08 & 3.55 & 0.60 & 2.04 & 0.02 & 4.69\\
J033233.48-275030.0 & 4.900 &C &abs. & 25.76 & 4.07 & 0.07 & 2.38 & 0.66 & 2.65\\
J033223.99-274107.9 & 4.920 &C &abs. & 25.26 & 2.38 & 0.98 & 2.31 & 0.75 & 4.51\\
J033247.66-275105.0~\tablenotemark{\star} & 4.920 &C &abs. & 25.55 & 2.57 & 0.98 & 2.35 & 1.01 & 2.66\\
J033234.49-274403.0 & 4.948 &C &em. & 26.04 & 3.45 & 0.51 & 1.49 &-0.08 & 9.19\\
J033225.32-274530.9 & 4.992 &B &em. & 26.70 & 4.45 & 0.11 & 2.69 & 0.53 & 0.55\\
J033221.30-274051.2 & 5.292 &A &em. & 25.23 & 5.36 & 0.10 & 2.01 & 0.57 & 3.58\\
J033245.43-275438.5 & 5.375 &A &em. & 25.15 & 6.08 & 0.03 & 2.86 & 0.79 & 1.94\\
J033224.40-275009.9 & 5.500 &C &abs. & 25.29 & 7.76 & 0.02 & 2.74 & 1.18 &-0.07\\
J033237.63-275022.4 & 5.518 &A &em. & 25.76 & 8.05 & 0.01 & 2.58 & 1.05 & 0.80\\
J033218.92-275302.7 & 5.563 &A &em. & 24.58 & 3.37 & 0.83 & 2.43 & 0.59 & 6.28\\
J033211.93-274157.1 & 5.578 &B &em. & 26.53 & 4.10 & 0.10 & 2.07 & 1.03 & 1.40\\
J033245.23-274909.9 & 5.583 &B &em. & 25.81 & 6.97 & 0.01 & 2.73 & 1.02 & 0.10\\
J033214.74-274758.7 & 5.939 &B &em. & 26.36 & 4.15 & 0.17 & 2.36 & 1.12 &-0.17\\
\tableline
\enddata
\tablenotetext{a}{Identify with \Lya\ and NV1240\AA~(and possibly CIV 1549\AA), QSO.}
\tablenotetext{\star}{For this source, the QF has been changed (B to C) to respect the online 
release of \cite{vanz08}, after re-analyzing the whole LBG sample.}
\end{deluxetable}

%%%%%%%%%%%%%%%%%%%%%%%%%
\clearpage
\begin{deluxetable}{lcccccccc}
\tabletypesize{\scriptsize}
\tablecaption{The spectroscopic sample of the \wi--band dropouts. Cloumns as described in
Table~\ref{tab:Bdrops}.\label{tab:Idrops}}
\tablewidth{0pt}
\tablehead{
\colhead{GOODS ID} & \colhead{$z$} & \colhead{QF} & 
\colhead{class} & \colhead{\wz\ } & \colhead{h.l.r.} & 
\colhead{S/G} & \colhead{(\wi\--\wz\ )} 
& \colhead{$(S/N)_{i}$}
}
\startdata
J033219.23-274545.5 & 0.000 &C &star & 23.47 & 2.64 & 0.98 & 1.35 & 37.91\\
J033218.19-274746.6 & 0.000 &B &star & 23.76 & 2.66 & 0.99 & 1.48 & 28.00\\
J033224.79-274912.9 & 0.000 &C &star & 24.95 & 2.77 & 0.99 & 1.60 & 10.92\\
J033238.80-274953.7 & 0.000 &C &star & 25.16 & 3.81 & 0.91 & 3.87 & 0.99\\
J033222.47-275047.4 & 0.000 &C &star & 24.42 & 2.70 & 0.99 & 1.74 & 15.04\\
J033238.02-274908.4 & 0.000 &B &abs. & 25.41 & 2.59 & 0.99 & 1.37 & 9.77\\
J033239.03-275223.1~\tablenotemark{\star\star} & 5.559 &C &em. & 25.72 & 4.08 & 0.18 & 1.66 & 5.16\\
J033215.90-274123.9 & 5.574 &B &em. & 25.48 & 8.31 & 0.00 & 1.51 & 4.56\\
J033227.91-274942.0 & 5.757 &C &em. & 26.91 & 3.48 & 0.27 & 1.60 & 2.45\\
J033255.32-275315.6 & 5.764 &B &em. & 26.15 & 8.21 & 0.01 & 1.41 & 3.53\\
J033225.61-275548.7~\tablenotemark{a} & 5.786 &A &em. & 24.69 & 3.69 & 0.64 & 1.65 & 11.87\\
J033246.04-274929.7~\tablenotemark{b} & 5.787 &A &em. & 26.11 & 4.12 & 0.02 & 1.92 & 3.06\\
J033254.10-274915.9 & 5.793 &C &em. & 25.26 & 10.84 & 0.00 & 1.90 & 2.52\\
J033240.01-274815.0 & 5.828 &A &em. & 25.34 & 3.95 & 0.39 & 1.47 & 7.92\\
J033233.19-273949.1~\tablenotemark{\star} & 5.830 &B &abs. & 25.41 & 6.74 & 0.01 & 2.18 & 3.30\\
J033249.98-274656.2 & 5.890 &B &em. & 26.25 & 3.47 & 0.50 & 1.65 & 3.76\\
J033224.97-275613.7 & 5.899 &B &em. & 26.78 & 4.10 & 0.07 & 1.99 & 1.75\\
J033239.06-274538.7 & 5.920 &B &em. & 27.05 & 3.80 & 0.90 & 1.53 & 2.45\\
J033228.19-274818.7 & 5.940 &B &em. & 26.48 & 3.97 & 0.18 & 1.59 & 3.08\\
J033215.76-274817.2 & 5.944 &C &em. & 26.09 & 5.08 & 0.17 & 1.75 & 3.01\\
J033236.47-274641.4~\tablenotemark{d} & 5.950 &C &abs. & 26.30 & 4.69 & 0.01 & 2.88 & 0.58\\
J033232.46-274001.9 & 5.977 &B &em. & 26.51 & 3.59 & 0.28 & 2.36 & 1.51\\
J033218.08-274113.1 & 5.979 &C &em. & 26.72 & 5.61 & 0.00 & 2.03 & 0.27\\
J033224.80-274758.8 & 5.996 &B &em. & 26.06 & 5.64 & 0.03 & 2.00 & 2.33\\
J033229.33-274014.3 & 6.000 &C &abs. & 26.81 & 2.77 & 0.90 & 2.12 & 1.90\\
J033246.43-275524.4 & 6.082 &B &em. & 26.80 & 4.68 & 0.75 & 2.35 & 0.15\\
J033223.84-275511.6 & 6.095 &B &em. & 26.31 & 3.08 & 0.67 & 3.22 &-0.27\\
J033229.84-275233.2 & 6.197 &B &em. & 26.26 & 7.52 & 0.25 & 2.48 & 1.16\\
J033222.28-275257.2 & 6.200 &C &abs. & 25.86 & 6.31 & 0.27 & 2.73 &-0.01\\
J033217.81-275441.6 & 6.277 &B &em. & 26.87 & 2.75 & 0.94 & 2.51 & 0.62\\
\tableline
\enddata
\tablenotetext{a}{also known as SBM03$\#$3 (\cite{bunk03}).}
\tablenotetext{b}{also known as GLARE$\#$3001 (\cite{stan04a}).}
\tablenotetext{c}{also known as SiD002 (\cite{dick04}), GLARE$\#$1042 (\cite{stan04a}), SBM03$\#$1, $\#$20104 (\cite{bunk04})}
\tablenotetext{d}{this has been identified by \cite{mal05} in the HUDF with ACS grism spectra.}\\  
\tablenotetext{\star}{For this source, the QF has been changed (C to B) to respect the online 
release of \cite{vanz08}, after re-analyzing the whole LBG sample.}
\tablenotetext{\star\star}{Redshift has been added to the previous LBG list (Vanzella et al. \ 2008), after re-analyzing the whole LBG sample and tacking together the unconclusive single spectra.}
\end{deluxetable}

%%%%%%%%%%%%%%%%% TABLE 2
\clearpage
\begin{table} 
\centering 
\caption{Fraction of confirmed dropout candidates, ``Nobs.'' indicates the
number of candidates observed. The number of confirmed high and low-redshift 
galaxies is reported in columns three and four, respectively
(with the fraction of ``em.'', ``abs.'' and ``comp.'' classes and
the fraction of QFs ``A'', ``B'' and ``C'').  
In columns 5 and 6 the average and
standard deviation of the redshift distribution for the confirmed high-z
sample are shown. In column 7 the completeness is reported. \label{tab:high-z}}
\begin{tabular}{llll|ccc} 
 \tableline  \tableline
classes &    Nobs.   & high-z                  &             low-z & measured\tablenotemark{a}          & expected\tablenotemark{a,b} & compl.\tablenotemark{a} \\
        &            & N $_{(A,B,C)}^{(em,abs,comp)}$   & N $_{(A,B,C)}^{(em,abs,comp)}$   &$<z>$$\pm$$\sigma$ &$<z>$$\pm$$\sigma$ &   \\
\tableline
  \wb--drop & 85   & 46 $_{~(27,~11,~8)}^{~(15,~21,~10)}$   & 2$_{~(0,~2,~0)}^{~(1,~1,~0)}$ &3.76$\pm$0.33 & 3.78$\pm$0.34 &5$\%$ \cr 
\\
  \wv--drop & 52   & 32 $_{~(9,~12,~11)}^{~(19,~13,~0)}$     & 4$_{~(2,~2,~0)}^{~(0,~4,~0)}$ &4.96$\pm$0.38 & 4.92$\pm$0.33 &14$\%$ \cr
\\
  \wi--drop & 65   & 28 $_{~(3,~13,~12)}^{~(24,~4,~0)}$      & 6$_{~(0,~2,~4)}^{~(0,~6,~0)}$  &5.90$\pm$0.18 & 5.74$\pm$0.36 & 29$\%$ \cr 
\tableline
  Fillers  & -    & 3 $_{~(1,~1,~1)}^{~(1,~2,~0)}$        & -                          & $3.4<z<5.5$               & - &- \cr
\\
  Serend.  & -    & 5 $_{~(0,~1,~4)}^{~(4,~1,~0)}$         &  -                         & $3.2<z<5.8$              & - &- \cr 
\tableline
 Sum          &202   & 114                                 & 12                         &              &               &       \cr
\end{tabular} 
\tablenotetext{a}{calculated down to \wz\ = 26.5}
\tablenotetext{b}{see \cite{giava04b}}
\end{table}

%%%%%%%%%%%%%%%%%%%%%%%%% 
\clearpage
\begin{deluxetable}{lcccccl}
\tabletypesize{\scriptsize}
\tablecaption{The spectroscopic sample of galaxies identify at redshift beyond 
3 serendipitously discovered and/or selected from the previous version
(v1.0) of the ACS catalogs and not satisfying the v2.0 one (see text for details). 
\label{tab:99drops}}
\tablewidth{0pt}
\tablehead{
\colhead{GOODS ID} & \colhead{$z$} & \colhead{QF} & 
\colhead{class} & \colhead{\wz\ } & 
\colhead{S/G} & \colhead{Comment}
}
\startdata
J033219.41-274728.4 & 3.250 &C &abs. & 24.65 &  0.33&serend. (B-V=1.30,V-z=0.48, v2.0)\\
J033234.40-274124.3 & 3.418 &B &abs. & 24.26 &  0.45&filler. (B-V=1.42,V-z=0.55, v2.0)\\
J033251.81-275236.5 & 3.468 &A &abs. & 25.03 &  0.04&filler. (B-V=1.15,V-z=0.36, v2.0)\\
J033206.53-274259.1 & 3.605 &C &abs. & 24.44 &  0.03&\wb--drop, from v1.0 (B-V=1.59,V-z=0.52, v2.0)\\
J033217.00-274113.7 & 4.414 &B &abs. & 25.09 &  0.02&\wv--drop, from v1.0 (B-V=1.90,V-z=1.67, v2.0)\\
J033218.27-274712.0 & 4.783 &C &em. & 27.60 &   0.90&serend. (close to \wv--drop, pentagon in Fig.~\ref{Vdrop}).\\
J033243.16-275034.6~\tablenotemark{a} & 4.838 &C &em. &- &- &serend. (See Fig. 2 of \cite{vanz06})\\
J033228.94-274128.2~\tablenotemark{a} & 4.882 &B &em. &- &- &serend. (See Fig. 13 of \cite{vanz05})\\
J033222.71-275154.4 & 4.900 &C &abs. & 25.55 &  0.02&\wv--drop, from v1.0 (V-i=1.76,i-z=0.30, v2.0)\\
J033249.15-275022.5 & 4.910 &C &abs. & 25.54 &  0.99&\wv--drop, from v1.0 (V-i=1.95,i-z=0.67, v2.0)\\
J033211.71-274149.6 & 4.912 &C &em. & 25.36 &  0.05&\wv--drop, from v1.0 (V-i=1.68,i-z=0.25, v2.0)\\
J033222.89-274521.0~\tablenotemark{b} & 5.128 &C &em. &- &- &serend.(LAE?)\\
J033216.55-274103.2 & 5.250 &C &abs. & 25.69 &  0.00&\wi--drop, from v1.0 (V-i=1.91,i-z=1.11, v2.0))\\
J033228.55-275621.8 & 5.492 &B &em. & 27.45 &  0.96&\wi--drop, from v1.0 (V-i=1.44,i-z=1.30, v2.0)\\
J033239.82-275258.1 & 5.543 &C &em. & 26.53 &  0.15&filler, (V-i=1.66,i-z=0.96, v2.0)\\
J033233.52-275532.2~\tablenotemark{c} & 5.740 &C &em. &- &- &\wi--drop, from v1.0.\\
J033201.96-274406.5 & 5.821 &C &em. & 26.19 & 0.01&\wi--drop, from v1.0 (V-i=1.49,i-z=1.20, v2.0)\\
\tableline
\enddata
\tablenotetext{a}{Sources not detected in the \wz\ band because blended to bright ones.}
\tablenotetext{b}{Source not detected in the \wz\ band, only visible in the \wi\ band.}
\tablenotetext{c}{Source originally detected in the v1.0 catalog, but not detected in the v2.0.}
\end{deluxetable}

%%%%%%%%%%%%%%%%%%%%
\clearpage
\begin{deluxetable}{lccc}
\tablecaption{Basic morphological parameters for \wb--band dropouts galaxies, dividing between
emitters and non-emitters. For the first four rows the values are reported in pixels, while the
Gini coefficient measure the nucleation of the source light (see text for details). \label{morph}}
\tablewidth{0pt}
\tablehead{
\colhead{} & \colhead{EM.} & \colhead{ABS.}
}
\startdata
             &        ($<z>$=3.757)      &   ($<z>$=3.735) \\
\tableline
a            &       $4.54\pm1.03$       &  $6.51\pm2.18$  \\
h.l.r.       &       $5.38\pm1.65$       &  $7.49\pm2.85$  \\
area         &       $305\pm117$         &  $438\pm172$    \\
FWHM         &       $10.38\pm4.06$      &  $20.01\pm11.97$ \\
Gini         &       $0.41^{+0.11}_{-0.06}$      &  $0.26^{+0.18}_{-0.10}$ \\ 
\enddata
\end{deluxetable}

%%%%%%%%%%%%%%%%% 
\clearpage
%

%%%%%%%%%% 
\clearpage
%

%%%%%%%%%%%%%%%%%  
%

%

%
%%%%%%%%%%%%%%%%%%%% 
\clearpage
%

%%%%%%%%%%%%%%%%%% 
\clearpage

%%%%%%%%%%%%%%%%%
\clearpage
%

%%%%%%%%%%%%%%%%%%% 
\clearpage
%

%%%%%%%%%%%%%%%%%%%%
\clearpage
%

%%%%%%%%%%%%%%%%%%%%%
\clearpage
%

%%%%%%%%%%%%%%%%%%%
\clearpage
%

%%%%%%%%%%%%%%%%%
\clearpage

%%%%%%%%%%%%%%%%%%%%
\clearpage
%

%%%%%%%%%%%%%%%% 
\clearpage
%

%%%%%%%%%%%%%%%% 
\clearpage
%

%%%%%%%%%%%%%%%% 
\clearpage
%

%%%%%%%%%%%%%%  
\clearpage
%

%%%%%%%%%%%%%%%%%%
\clearpage
%

%%%%%%%%%%%%%%%%%%
\clearpage

\end{document}